\documentclass[twocolumn,preprintnumbers,amsmath,amssymb,superscriptaddress]{revtex4}
\usepackage{graphicx}
\usepackage{dcolumn}
\usepackage{bm}
\usepackage{soul}
\usepackage{color}
\usepackage{epstopdf}
\usepackage[version=3]{mhchem}
\usepackage{lipsum}
\usepackage[outercaption]{sidecap}
\usepackage{floatrow}
\usepackage{hyperref}

\begin{document}

\title{Influence of Longer Range Transfer of Vapor Interface Modified Caging Constraints on the Spatially Heterogeneous Dynamics of Glass-Forming Liquids}
\affiliation{Departments of Physics, $^2$Materials Science, $^3$Chemistry and $^4$Chemical $\&$ Biomolecular Engineering, $^5$Materials Research Laboratory, University of Illinois, 1304 West Green Street, Urbana, IL 61801\\
$^6$Faculty of Materials Science and Engineering, Phenikaa Institute for Advanced Study, Phenikaa University, Hanoi 100000, Vietnam\\
$^7$Faculty of Information Technology, Artificial Intelligence Laboratory, Phenikaa University, Hanoi 100000, Vietnam}
\author{Anh D. Phan$^{1,5-7}$}
\email{anh.phanduc@phenikaa-uni.edu.vn}
\author{Kenneth S. Schweizer$^{1-5}$}
\email{kschweiz@illinois.edu}
\date{\today}

\date{\today}

\begin{abstract}
Based on the Elastically Collective Nonlinear Langevin Equation (ECNLE) theory of bulk relaxation in glass-forming liquids, and our recent ideas of how interface-nucleated modification of caging constraints are spatially transferred into the interior of a thick film, we present a force-based theory for dynamical gradients in thick films with one vapor interface that includes collective elasticity effects. Quantitative applications to the foundational hard sphere fluid and polymer melts of diverse fragilities are presented. We predict a roughly exponential spatial variation of the total activation barrier which is non-perturbatively modified to $\sim 10$ particle diameters into the film. This leads, to leading order, to the prediction of a reduced alpha relaxation time gradient of a double exponential form characterized by a nearly constant spatial decay length, in qualitative accord with simulations. The relaxation acceleration at the surface grows exponentially with increasing packing fraction or decreasing temperature. The rate of increase with cooling of the alpha time strongly weakens upon approaching the vapor interface, with the top two layers remaining liquid-like down to $\sim 80-85\%$ of the bulk glass transition temperature. These spatially heterogeneous changes of the temperature variation of the alpha time result in a large and long range gradient of the local glass transition temperature. An average interfacial layer thickness relevant to ensemble-averaged dielectric and other experiments is also computed. It is predicted to be rather large at the bulk glass transition temperature, decreasing roughly linearly with heating. Remarkably, to leading order the ratio of this layer-averaged interfacial relaxation time to its bulk analog is, to leading order, invariant to chemistry, volume fraction and temperature and of modest absolute magnitude. Spatially inhomogeneous power law decoupling of the alpha relaxation time from its bulk value is predicted, with an effective exponent that decays to zero with distance from the free surface in a nearly exponential manner, trends which are in qualitative accord with recent simulations. This behavior and the double exponential alpha time gradient are related, and can be viewed as consequences of an effective quasi-universal factorization of the total barrier in films into the product of its bulk temperature dependent value times a function solely of location in the film.
\end{abstract}

\maketitle


\section{Introduction}
Activated dynamics, mechanical properties and vitrification in thin films of glass-forming liquids is a problem of great scientific interest \cite{1,2,3,4,5}, which is also important in many materials applications \cite{6,7,8,9}. Despite intense experimental, simulation and theoretical efforts over the past two decades \cite{2,5,10,11,12,13,14,15,16,17,18}, the key physical mechanisms underlying the observed phenomena remain intensely debated. This situation reflects the complexity of activated structural relaxation in bulk liquids10 and the additional large complications of geometric confinement, interfaces and spatial inhomogeneity. 

A quantitative force-level statistical mechanical approach for structural (alpha) relaxation in bulk colloidal, molecular and polymer liquids, the “Elastically Collective Nonlinear Langevin Equation” (ECNLE) theory \cite{19,20,21,22,23,24}, has been recently developed and widely applied to real materials. Structural relaxation is described as a coupled local-nonlocal activated process involving large amplitude cage-scale hopping facilitated by a small amplitude longer-range collective elastic deformation of the surrounding liquid. Quantitative tractability for real world liquids is achieved based on an a priori mapping of chemical complexity to a thermodynamic-state-dependent effective hard sphere fluid \cite{20,23}.

Very recently, we formulated a new theory for how caging constraints in glass-forming liquids at a solid or vapor surface of a \text{thick} film are modified and spatially transferred into its interior in the context of the dynamic free energy concept of NLE theory \cite{25}. The caging component of the dynamic free energy varies very nearly exponentially with distance from the interface, saturating deep enough into the film with a correlation length of modest size and weak sensitivity to thermodynamic state. This imparts a roughly exponential spatial variation of dynamical quantities such as the transient localization length, jump distance, and cage barrier. The theory was implemented for the hard sphere fluid and diverse interfaces: vapor, rough pinned particle solid, vibrating (softened) pinned particle solid, and smooth hard wall. The crucial difference between these systems arises from the first layer where dynamical caging constraints can be either weakened, softened, or hardly changed depending on the microscopic boundary conditions. Numerical calculations established the spatial dependence and fluid volume fraction sensitivity of \textit{local} dynamical property gradients for five different model interfaces. 

Free standing thin films with two vapor interfaces, or semi-infinite thick films with one vapor surface, are the simplest realizations of confined anisotropic systems. Extensive experimental \cite{1,2,3,25,26,27,28,29,30,31,32,33,34} and simulation \cite{2,5,14,35,36,37,38} efforts mainly on polymer films suggest a spatially inhomogeneous and very large speed up of relaxation with mobile layers extending far from the interface resulting in large film-averaged reductions of $T_g$. Conceptual puzzles include how mobility changes nucleated at a vapor interface "propagate" into the film, the physical origin of the simulation finding that the relaxation time gradient for free standing substrate films appears to have (to leading order) a "double exponential" form \cite{13,14,37,39,40,41,42}, and the very recent simulation finding by Simmons and co-workers of so-called "decoupling" of the temperature and film location dependences of the alpha time gradient characterized with an effective scaling exponent \textit{function} that varies exponentially with distance from the interface \cite{41}.

The present article addresses the above open questions and related issues by applying our new theory of spatial mobility transfer in a thick film with one vapor interface within the full ECNLE framework \cite{25} that includes the collective elasticity effects. We focus on representative polymer liquids, but the findings are broadly relevant to molecular liquids and colloidal suspensions. The basic theoretical ideas are reviewed in section II.  Sections III-VI then quantitatively study the foundational hard sphere fluid and specific polymers of diverse bulk dynamic fragilities. Results are presented for the total and collective elastic barrier spatial gradients, the alpha time gradient and origin of its double exponential nature, the amplitude and decay length of this dynamical gradient, spatial decoupling of the relaxation time and how the effective exponent varies with location in the film, a mean interfacial layer width and corresponding relaxation time, and the magnitude and functional form of the $T_g$ gradient. The article concludes with a summary in section VII. The Appendix discusses technical details of our comparison of theory and simulation.

\section{Background: ECNLE Theory of Bulk Liquids and Films}
ECNLE theory in bulk liquids and under confined conditions has been discussed in great depth in a series of previous papers \cite{19,20,21,22,23,24,43,44,45}. Here we concisely recall the key elements. 
\subsection{Bulk Liquids}
ECNLE theory describes tagged particle activated relaxation as a mixed local-nonlocal hopping event \cite{19,20,21,22,23,24}. Figure 1 shows a schematic of the physical elements in the context of a liquid of spheres of diameter, $d$, at a packing or volume fraction $\Phi$. The foundational dynamical quantity is the angularly-averaged instantaneous displacement-dependent dynamic free energy, $F_{dyn}(r)=F_{ideal}(r)+F_{caging}(r)$, where $r$ is the displacement of a particle from its initial position. The localizing “caging” contribution $F_{caging}(r)$ is constructed \textit{solely} from knowledge of $\Phi$ and the pair correlation function $g(r)$ or structure factor $S(q)$, where $q$ is the wavevector, computed here using Percus-Yevick (PY) integral equation theory. It captures kinetic constraints on the nearest neighbor cage scale defined from the location of the first minimum of $g(r)$ ($r_{cage}\approx 1.5d$). Key local length and energy scales (see Figure 1) are the minimum and maximum of the dynamic free energy ($r_L$ and  $r_B$), corresponding harmonic curvatures ($K_0$ and $K_B$), jump distance $\Delta{r}=r_B-r_L$, and local cage barrier height, $F_B$.  

\begin{figure}[htp]
\includegraphics[width=8.5 cm]{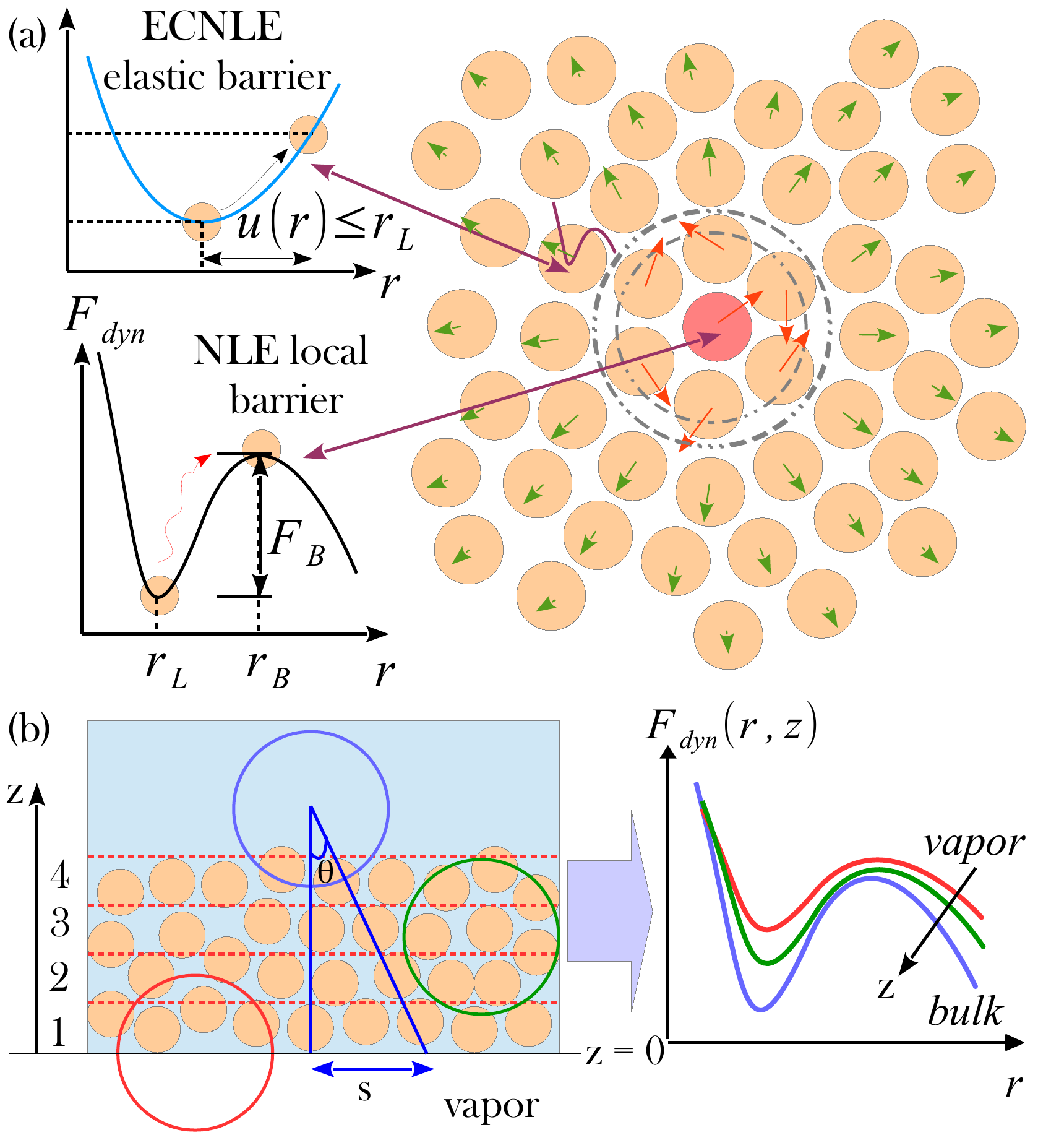}
\caption{\label{fig:1}(Color online) (a) Schematic of the fundamental relaxation event for a bulk dense fluid of spheres (Kuhn segments for polymers) which involves large amplitude cage-scale hopping and a spatially longer-range collective elastic harmonic motion of particles outside the cage. Hopping trajectories are described by the dynamic free energy as a function of particle displacement, and the jump distance sets the amplitude of the elastic displacement field. Various key length and energy scales are indicated. (b) Schematic illustration of the layer-like model employed to describe the surface nucleated change of caging constraints and how the dynamic free-energy is modified near a vapor interface as a function of the distance from it, $z$.}
\end{figure}

In the spirit of phenomenological “elastic models” \cite{46,47}, large amplitude local hopping is argued to be strongly and mechanistically coupled to a longer range collective elastic adjustment of all particles outside the cage needed to create the extra space to allow a hop \cite{21}. The form of the radially-symmetric elastic displacement field (Figure 1) is calculated for $r>r_{cage}$ based on a continuum mechanics analysis \cite{46} with a microscopically-determined amplitude set by the cage expansion length, $\Delta{r_{eff}}$ \cite{19}:
\begin{eqnarray}
u(r)=\Delta{r_{eff}}\left(\frac{r_{cage}}{r}\right)^2,\;\Delta{r_{eff}}\approx\frac{3\Delta{r^2}}{32r_{cage}},\; r>r_{cage}
\label{eq:1}
\end{eqnarray}
where $\Delta{r}\approx 0.2-0.4d$ and grows with density and cooling. The elastic barrier is computed microscopically by summing over all harmonic particle displacements outside the cage region \cite{19}: 
\begin{eqnarray}
F_{elastic}&=&\rho(K_0/2)\int_{r_{cage}}^{\infty}dr4\pi r^2u^2(r)g(r)\nonumber\\
&=&12\Phi\left(\frac{r_{cage}}{d}\right)^3\Delta r_{eff}^2K_0
\label{eq:2}
\end{eqnarray}
where $r$ is relative to the cage center. The sum of the coupled local and elastic collective barriers determine the mean total barrier for the alpha relaxation process: 
\begin{eqnarray}
F_{total}=F_B+F_{elastic}
\label{eq:3}
\end{eqnarray}
The elastic barrier increases much more strongly with increasing density or cooling than its cage analog, and dominates the rate of alpha time growth for fragile liquids as the laboratory $T_g$  is approached \cite{19,20,21,22,23,24}. 

A generic measure of the average structural relaxation time follows from a Kramers calculation of the mean first passage time for barrier crossing. For barriers in excess of a few $k_BT$ one has \cite{19,20}:   
\begin{eqnarray}
\frac{\tau_\alpha}{\tau_s}=1+\frac{2\pi}{\sqrt{K_0K_B}}\frac{k_BT}{d^2}\exp\left(\frac{F_B+F_{elastic}}{k_BT}\right)   		
\label{eq:4}
\end{eqnarray}
The alpha time is expressed in units of a "short time/length scale" relaxation process as discussed elsewhere \cite{19,20,21,22,23,24}. Physically, it captures the alpha process in the absence of strong caging defined by the parameter regime where no barrier is predicted by NLE theory (e.g., $\Phi<0.43$ for hard spheres \cite{45}). 
\subsection{Mapping for Molecular and Polymeric Liquids}
The theory is rendered quantitatively predictive and  \textit{quasi-universal} for molecular liquids via a mapping \cite{19,20,21,22,23,24} to an effective hard sphere fluid guided by the requirement it \textit{exactly} reproduces the equilibrium dimensionless density fluctuation amplitude (compressibility) of the experimental liquid of interest \cite{48}, $S_0(T)=\rho k_BT\kappa_T$. Using PY theory \cite{48}, the mapping is \cite{20,21,22,23,24}:  
\begin{eqnarray}
S_{0}^{HS}&=&\frac{\left(1-\Phi_{eff}\right)^4}{\left(1+2\Phi_{eff}\right)^2}=S_{0,expt}=\rho_sk_BT\kappa_T\nonumber\\
&\approx&N_{s}^{-1}\left(-A^{*}+\frac{B^{*}}{T}\right)^{-2}
\label{eq:5}
\end{eqnarray}
The final equality describes well experimental $S_{0,expt}(T)$ data of molecular and polymeric liquids. This mapping determines a material-specific, temperature-dependent effective hard sphere packing fraction, $\Phi_{eff}(T)$. Known chemically-specific parameters enter \cite{20,21,22,23,24}: $A^{*}$ and $B^{*}$ (interaction site level entropic and cohesive energy equation-of-state (EOS) parameters, respectively), the number of elementary sites that define a rigid molecule, $N_s$ (e.g., $N_s=18$ for orthoterphenyl, C$_{18}$H$_{18}$ (OTP)), and hard sphere diameter, $d$. With this mapping, ECNLE theory can make mean alpha time predictions with no adjustable parameters, and accurately captures relaxation time data over 12-14 decades for nonpolar organic molecules \cite{20,21}. 

For polymers, the liquid is modeled as a fluid of disconnected spheres of size determined by the Kuhn length and its space filling volume \cite{23}. Recently it was argued a modest degree of \textit{nonuniversality} associated with how conformational isomerism and other chemical complexities on the sub-Kuhn segment scale influence the jump distance is important \cite{24}. Specifically, one material-specific number, $\lambda$, was introduced that scales the jump distance as $\Delta{r}\rightarrow\lambda\Delta{r}$. This introduces a non-universality of the relative importance of collective elastic versus caging effects. Based on this elaboration, very good agreement between theory and experiments for $T_g$, fragility \textit{and} the full temperature dependence of the alpha time of chemically diverse polymers were simultaneously obtained \cite{24}. 

\begin{figure}[htp]
\includegraphics[width=8.5 cm]{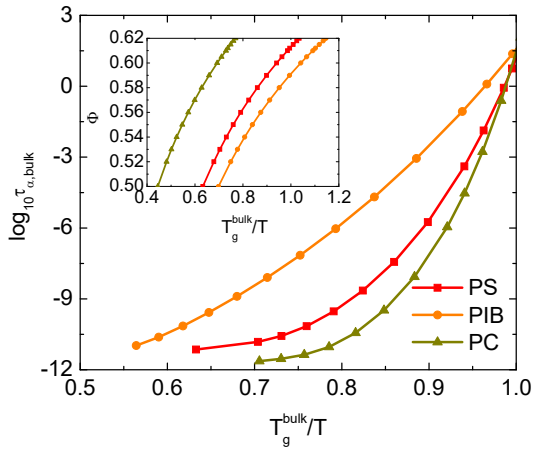}
\caption{\label{fig:2}(Color online) Main frame: Logarithm of the bulk alpha time (in seconds) as a function of inverse temperature normalized by the theoretical glass transition temperature for PS, PIB and PC. The predicted $T_g$ values for PS, PC and PIB are 430, 420, and 215 K, respectively \cite{44}. Inset: Effective volume fraction versus temperature normalized by the bulk glass transition temperature for the same 3 polymers.}
\end{figure}

In this article, we carry out quantitative applications to three representative polymers (the results are also representative of nonpolar molecules such as OTP) with known parameters of $N_s(PS)=38.4$, $d_{PS}=1.16$ nm, $A^{*}(PS)=0.618$, $B^{*}(PS)=1297$ K, $\lambda_{PS}=1$ for polystyrene (PS), $N_s(PC)=30.2$, $d_{PC}=1.04$ nm, $A^{*}(PC)=0.674$, $B^{*}(PC)=1280$ K, $\lambda_{PC}=\sqrt{2}$ for polycarbonate (PC), and $N_s(PIB)=13.6$, $d_{PIB}=0.89$ nm, $A^{*}(PIB)=0.627$, $B^{*}(PIB)=1353$ K, $\lambda_{PIB}\approx 0.47$ for polyisobutylene (PIB) \cite{24,44}. As previously shown \cite{20,44}, the effective volume fraction increases linearly (sub-linearly) with decreasing (increasing) temperature (inverse temperature) (see Figure 2). The main frame of Figure 2 shows predictions \cite{24} of the segmental relaxation time for PS, PC and PIB in the classic Angell representation. 
\subsection{Vapor Interface Thick Films}
A new theory for how a single vapor interface of a \textit{thick} film modifies the spatial dependence of caging constraints in the dynamic free energy was recently formulated \cite{25}. Its basic predictions for local dynamical quantities were worked out, but the consequences for the elastic barrier, alpha time gradient and many other properties were not analyzed. Here we briefly recall the new advance. 

We focused solely on \textit{purely} dynamical consequences of an interface, and hence the density and pair correlations in the film are taken to be spatially \textit{independent and identical} to those in the bulk. For a vapor interface, relaxation is sped up at the surface due to reduction of caging forces. For a tagged particle at the center of a cage, the dynamic free energy in layer $i$ is:
\begin{eqnarray}
F_{dyn}^{(1)}(r)=\frac{1}{2}F_{dyn}^{bulk}(r)+\frac{1}{2}F_{dyn}^{(i-1)}(r),\quad i\geq 1
\label{eq:6}
\end{eqnarray}
Per Figure 1b, the convention is $i=1$ corresponds to the first layer at $z=0$. Equation 6 implies an equal contribution of caging forces due to particles below and above the tagged particle. In a thick film, the upper half is taken to be unaffected by the vapor interface, and the caging component of the dynamic free energy is reduced in the first layer due to missing nearest neighbors. Equation 6 then yields:
\begin{eqnarray}
F_{dyn}^{(i)}(r)=F_{ideal}(r)+\left(1-\frac{1}{2^i}\right)F_{caging}^{bulk}(r)
\label{eq:7}
\end{eqnarray}
Importantly, the bulk form is recovered in an essentially exponential manner \cite{25} (since $2^{-i}=e^{-(z/d)\ln(2)}$) with a decay length of roughly $\sim 1.4d$.

A vapor-liquid interface also alters the elastic displacement field generated by cage expansion \cite{46}. We employ our most recent treatment of this problem \cite{44} based on a displacement field of the same form as in the bulk but modified by the molecular boundary condition of vanishing amplitude at the surface:
\begin{eqnarray}
u(r,\theta,z)=A_s(\theta,z)r+\frac{B_s(\theta,z)}{r^2}
\label{eq:8}
\end{eqnarray}
where $A_s(\theta,z)$ and $B_s(\theta,z)$ are chosen to enforce the boundary condition $u(r,\theta,z=0)=0$, and $\theta$ is defined in Figure 1b. At the cage surface $u(r,\theta,z=r_{cage})=\Delta{r_{eff}}$, which depends on location in the film via the spatially inhomogeneous dynamic free energy.  After straightforward calculation, we derived \cite{44}:
\begin{eqnarray}
A_s(\theta,z)=-\frac{\Delta{r_{eff}}r_{cage}^2\cos^3\theta}{z^3-r_{cage}^3\cos^3\theta}=-\frac{\Delta{r_{eff}}r_{cage}^2}{\left(z^2+s^2\right)^{3/2}-r_{cage}^3}\nonumber\\
B_s(\theta,z)=\frac{\Delta{r_{eff}}(z)r_{cage}^2z^3}{z^3-r_{cage}^3\cos^3\theta}=\frac{\Delta{r_{eff}}\left(z^2+s^2\right)^{3/2}}{\left(z^2+s^2\right)^{3/2}-r_{cage}^3}
\label{eq:9}
\end{eqnarray}
where the variable $s$ is defined in Figure1b. The collective elastic barrier then follows as an integral over the film particles at fixed distance, $z$, from the surface where the displacement field vanishes \cite{44}:
\begin{eqnarray}
F_{elastic}=\frac{3\Phi}{\pi d^3}\int_{V_{film}}d\vec{r}u^2(r,\theta,z)K_0(r,z)
\label{eq:10}
\end{eqnarray}

The inset of Figure 3 shows the collective elastic barrier gradient normalized by its bulk value for the hard sphere fluid. Importantly, its functional form is almost independent of packing fraction, as was demonstrated previously to be true of the local cage barrier \cite{25}. It is very strongly reduced close to the vapor surface by roughly an order of magnitude, and recovers (by eye) bulk behavior at $\sim 12-14$ particle diameters into the film. More quantitatively (as shown), its form is reasonably well fit by an exponential function with a decay length $\sim 3.5d$, as found previously \cite{25} for the local barrier, $F_B(z)$, albeit with a decay length is larger than its cage analog of $\sim 1.6d$. We emphasize that although the elastic barrier is associated with a “long range” collective elastic fluctuation, the amplitude of the displacement field and energy scale are determined by the cage-scale dynamic free energy. Specifically, $F_{elastic}(z)\propto\left(\Delta{r(z)}\right)^4K_0(z)\propto\left(\Delta{r(z)}\right)^4r_L^{-2}(z)$, where the jump distance and localization length both vary to a good approximation in an exponential manner \cite{25} with $z$. The exponential variation of the elastic barrier extends to $\sim 10d$ from the surface, but beyond that the asympototic approach to the bulk value varies inversely with $z$ due to the cutoff of the displacement field at the vapor interface \cite{44}. Very importantly, this power law cuttoff effect is now a rather minor contribution to the spatial variation of the elastic barrier, as opposed to prior studies \cite{44,49,50} which did not take into the new mobility transfer effects \cite{25}.   

\begin{figure}[htp]
\includegraphics[width=8.5 cm]{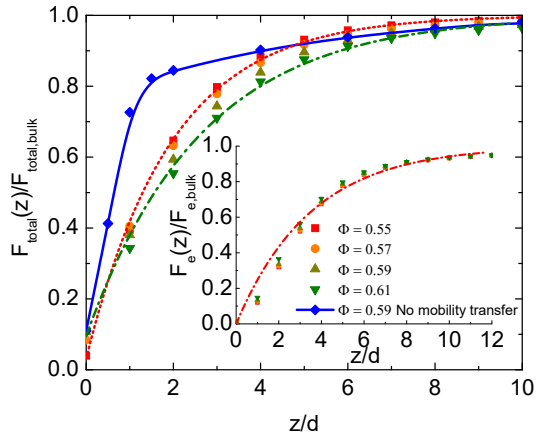}
\caption{\label{fig:3}(Color online) Normalized (to the bulk value) total barrier as a function of distance from the vapor interface for $\Phi=0.55, 0.57, 0.59, 0.61$ hard sphere fluids. Blue diamonds are calculations without mobility transfer effects (indicated as “no mobility transfer” in legend) at $\Phi=0.59$, and the curve through them is a guide to the eye. The dashed-dotted and dotted curves correspond to exponential fits, which for $\Phi=0.55$ and 0.61 are $1-0.968e^{-z/1.958d}$ (red) and $1-0.904e^{-z/2.644d}$ (green), respectively. Inset: normalized collective elastic barrier as a function of z at the same packing fractions  as in the main frame. For $\Phi=0.55$, $F_{elastic}(z)/F_{elastic}^{bulk}$ is fit by the dash-dot curve given by $1-e^{-z/3.516d}$.}
\end{figure}

The main frame of Figure 3 shows the analogous results for the \textit{total} dynamic barrier in the film normalized by its bulk liquid analog. This is the most fundamental and foundational result of this article. By eye, bulk behavior is recovered at $\sim 10$ particle diameters into the film. Although the local and collective elastic barriers both vary nearly \textit{exponentially} in space, they have different decay lengths. But, interestingly, Figure 3 shows that the theory predicts to a good approximation that the total barrier varies exponentially in space with an apparent decay length in between that of the cage and elastic barriers. Specifically, the latter is $\sim 2(2.6)d$ for the lower (higher) packing fractions. Hence, the total barrier ratio, $F_{total}(z,\Phi)/F_{total}^{bulk}(\Phi)$, has a similar $z$-dependence for all packing fractions, which is a crucial result of the present work. Also shown for one packing fraction is the analogous result if there is no mobility transfer effects \cite{44}. The barrier gradient is much shorter range and of a very different functional form.  

The above results are critical for all of our numerical results given below. When analyzing and physically interpreting these calculations, an "ideal factorization" scenario is considered motivated by our results discussed above that found to a good approximation:
\begin{eqnarray}
F_{total}(z,\Phi)/F_{total}^{bulk}(\Phi)=f(z)
\label{eq:11}
\end{eqnarray}
where $f(z)$ is, to leading order (Figure 3), an exponential function independent of thermodynamic state.

We now have all the information required to predict the most fundamental quantity - the alpha relaxation time spatial gradient. Below it is analyzed first for the foundational hard sphere fluid, and then for polymer liquids. Questions of interest include the functional form of the gradient, its amplitude at the surface and spatial range as a function of temperature, density and chemistry, other characteristic length scales relevant to simulation and experimental studies, the spatial dependence of the temperature variation of the alpha time (dynamic fragility), gradients of the local glass transition temperature, and inhomogeneous decoupling of film and bulk alpha relaxation times.
\section{Alpha Relaxation Time Gradients, Amplitudes, and Lengths Scales}
We first present numerical results for the alpha relaxation time spatial gradient which we view as the most fundamental physical quantity for understanding glassy dynamics near interfaces or in films. For nearly two decades it has been \textit{empirically} analyzed in simulation studies based on the 2-parameter “double exponential” form \cite{13,14,37,39,40,41,42}
\begin{eqnarray}
\ln\left(\frac{\tau_{\alpha,bulk}}{\tau_{(z)}}\right)=A(\Phi)\exp\left(-\frac{z}{\xi(\Phi)}\right)
\label{eq:12}
\end{eqnarray}
where the amplitude, $A(\Phi)$, at the surface and the penetration or decay length, $\xi(\Phi)$, are defined. The former is essentially the ratio of the dynamic barrier in the bulk to that \textit{at} the surface $(z=0)$. Importantly, prior simulations for vapor interfaces \cite{13,14,37,41} have found that $A$ grows strongly with cooling or increasing density, while the decay length is of modest size and weakly temperature-dependent. 
\subsection{Functional Form of the Alpha Time Spatial Gradient}
Figure 4a presents in a log-linear format calculations of the inverse alpha time normalized by its bulk analog as a function of distance from the vapor interface for the hard sphere fluid. Exponential fits are shown, which one can see are quite good for $\Phi\geq 0.57$. To the best of our knowledge, this provides the first potential theoretical basis for the so-called "double exponential" form of the normalized alpha time gradient. The amplitude factor quantifies a dramatic speed up of relaxation at the vapor surface that grows from a factor of $\sim 7$ to over $10$ decades as packing fraction increases. The bulk alpha time is recovered at a rather large distance from the surface which grows with packing fraction. The latter trend follows from the fact that although the \textit{normalized} barrier gradient decays to unity at a nearly universal reduced distance from the surface, its absolute value does not. The detailed behavior of the two gradient parameters are discussed in the next section. The inset of Figure 4a shows an even more discriminating double logarithmic plot of the same results. The double exponential form is not exact, with the expected deviations at very small and large distances from the interface, but it is a good description on intermediate length scales. Moreover, this form has the major practical advantage that it allows a continuous function to be discussed in terms of only 2 quantities of clear physical meaning: an surface amplitude factor and a decay length. 

\begin{figure}[htp]
\includegraphics[width=8.5 cm]{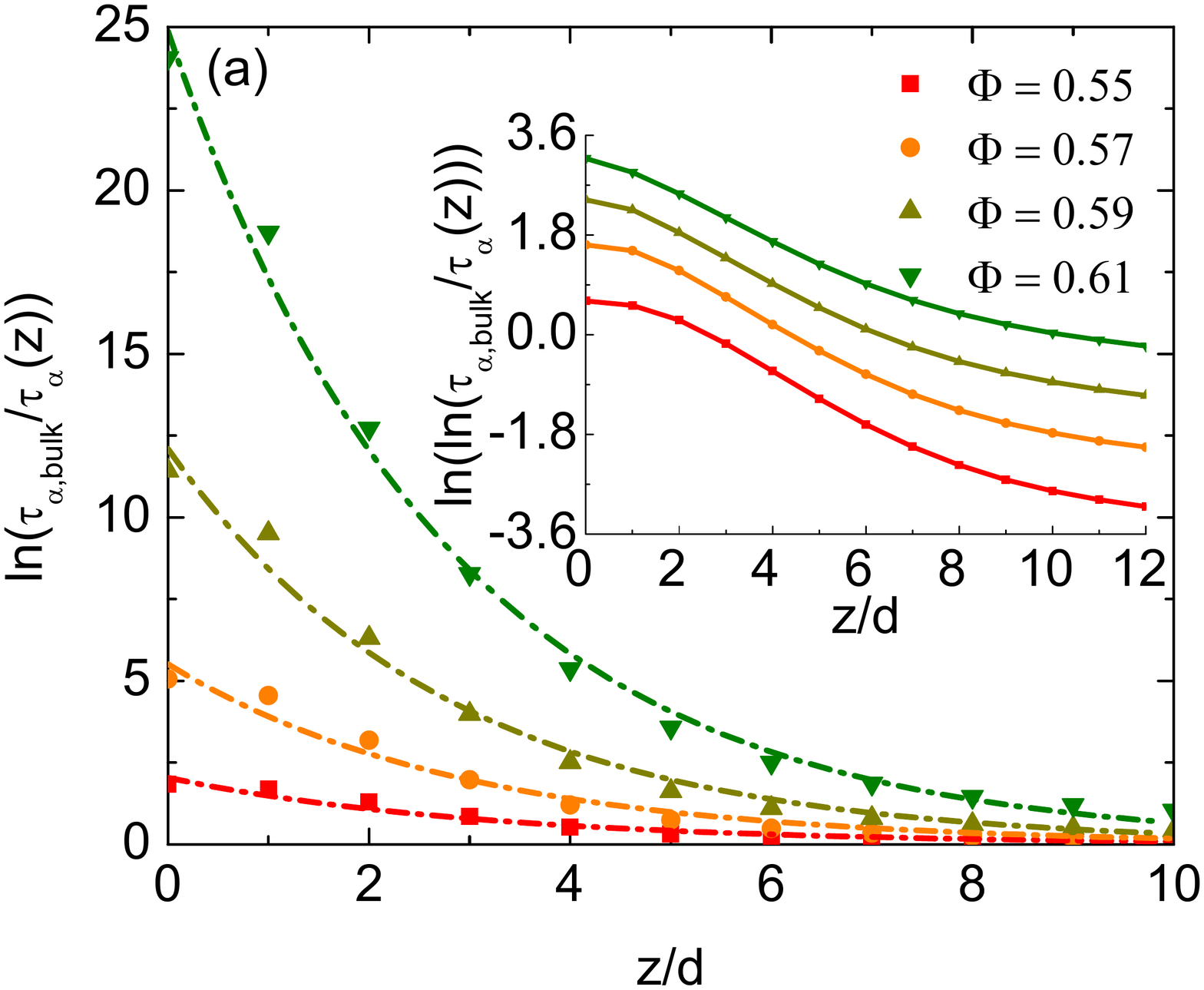}
\includegraphics[width=8.5 cm]{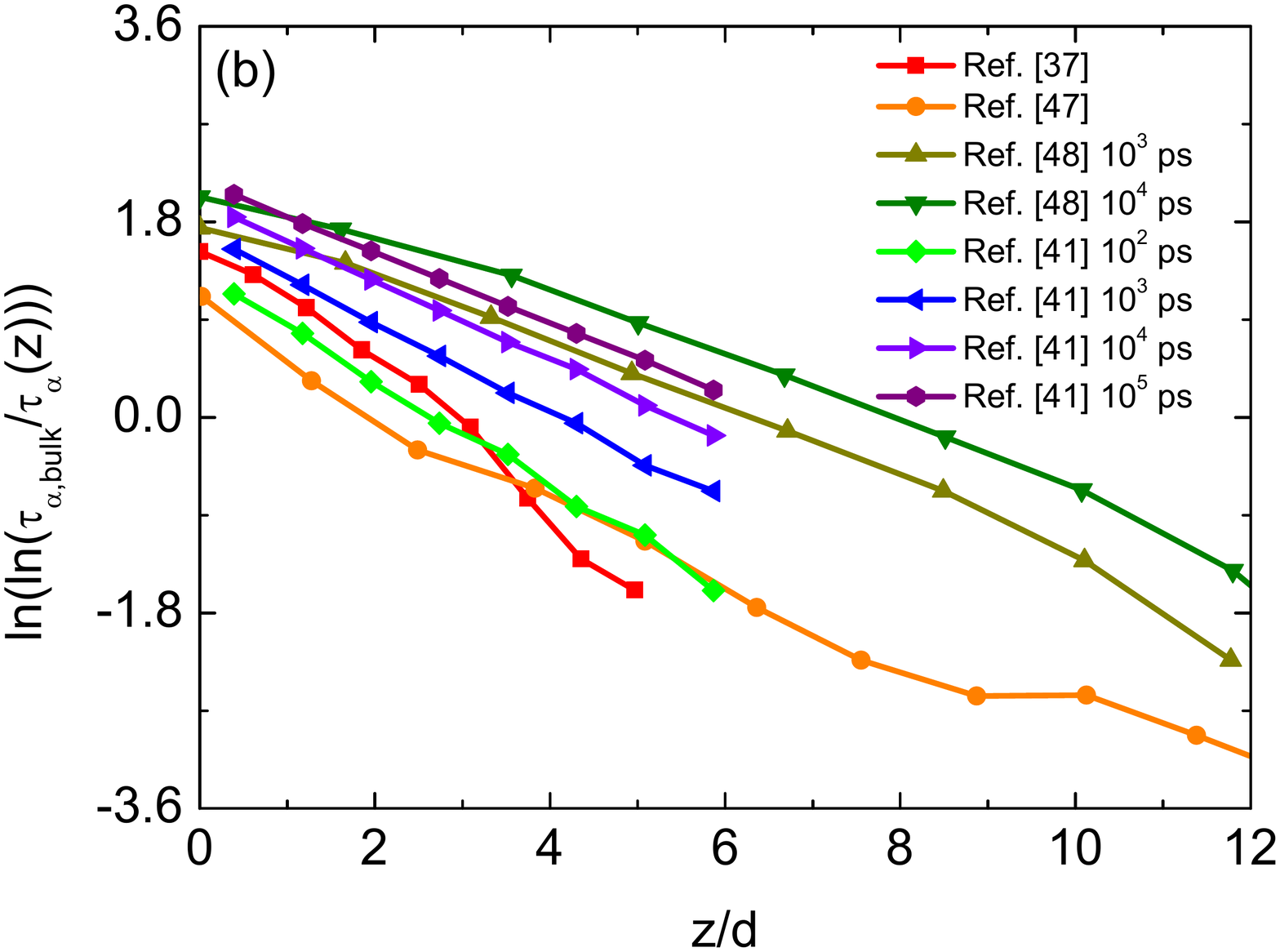}
\caption{\label{fig:4}(Color online) (a) Natural logarithm of the inverse alpha time normalized by its bulk value at a distance $z$ from the vapor surface. The dash-dot curves are exponential fits of the theoretical data points. For context, we note that for PS the volume fractions $\Phi=0.55, 0.57, 0.59, 0.61$ correspond to $T/T_g^{bulk}=1.32, 1.21, 1.11, 1.01$, respectively. Inset: Same results as main frame but plotted in a double natural logarithm format. Fits to the intermediate range of data using eq 12 yield length scales, $\xi$, of $2.3d, 2.45d, 2.55d, 2.7d$ for $\Phi=0.55, 0.57, 0.59, 0.61$, respectively. (b) Double natural logarithm of the inversely normalized alpha relaxation for the 4 simulation studies discussed in the text. State points for some simulations are indicated by the corresponding dimensionless bulk relaxation time. Simulation data for an atomistic PS model \cite{37} ($\xi=1.99d$), coarse grained PS model \cite{51} ($\xi=2.14d$), LJ coarse grained polymer model \cite{52} ($\xi=3.51d$ and $4.03d$ for $10^3$ and $10^4$ ps, respectively), and LJ bead-spring polymer model \cite{41} for free-standing films ($\xi=2.045d, 2.43d, 2.73d$ and $3.065d$ for $10^2, 10^3, 10+^4$, and $10^5$ ps, respectively). }
\end{figure}

Figure 4b shows a double logarithmic representation of the results of four different simulation studies - two based on atomistic \cite{37} or lightly coarse grained models \cite{51} of polystyrene melts, one based on a more heavily coarse-grained bead-spring polymer model \cite{52}, and one for an attractive bead-spring polymer model \cite{41}; see Appendix A for a more detailed description of the models. Recall simulations typically measure relaxation times only over 3-5 decades, which we estimate correspond to effective packing fractions in the range $\sim 0.55-0.58$. Despite the differences in simulation models, they all show basically the same double exponential form. Moreover, the curves are largely parallel for different temperatures indicating a very modest growth of the length scale in eq 12 with cooling, as also found in simulations of atomic models \cite{39,40,42}. The reduction of the alpha time at the surface in Figure 4b grows strongly with cooling by a factor that varies from $\sim 6$ to $\sim 1640$. Overall, these trends are consistent with the theory. At large distances from the interface, whether the noisy simulation data bends down or up in Figure 4b is unclear and perhaps model specific. However, the theory predicts the double logarithm representation of the inversely normalized relaxation times is roughly linear in the range $2d<z\leq 8d$, in qualitative accord with simulations \cite{37,39,40,41,42,51,52} to within their inevitable variability and uncertainties. Appendix A discusses quantitative issues that arise in our comparison of theory with simulation.

We note that although our earlier version of ECNLE theory for free standing films without the mobility transfer effect \cite{44} provided reasonable results for \textit{film-averaged} properties, a double exponential alpha time gradient was not predicted. This clearly demonstrates that the mobility transfer physics is crucial for determining the \textit{shape} and double-exponential form of the relaxation time gradient. This represents a third important physical process in ECNLE theory that coexists with the loss of neighbors very near the surface, and the cutoff of the collective elasticity displacement field at the vapor interface, which were analyzed in earlier theoretical work \cite{44,49,50}.

\begin{figure}[htp]
\includegraphics[width=8.5 cm]{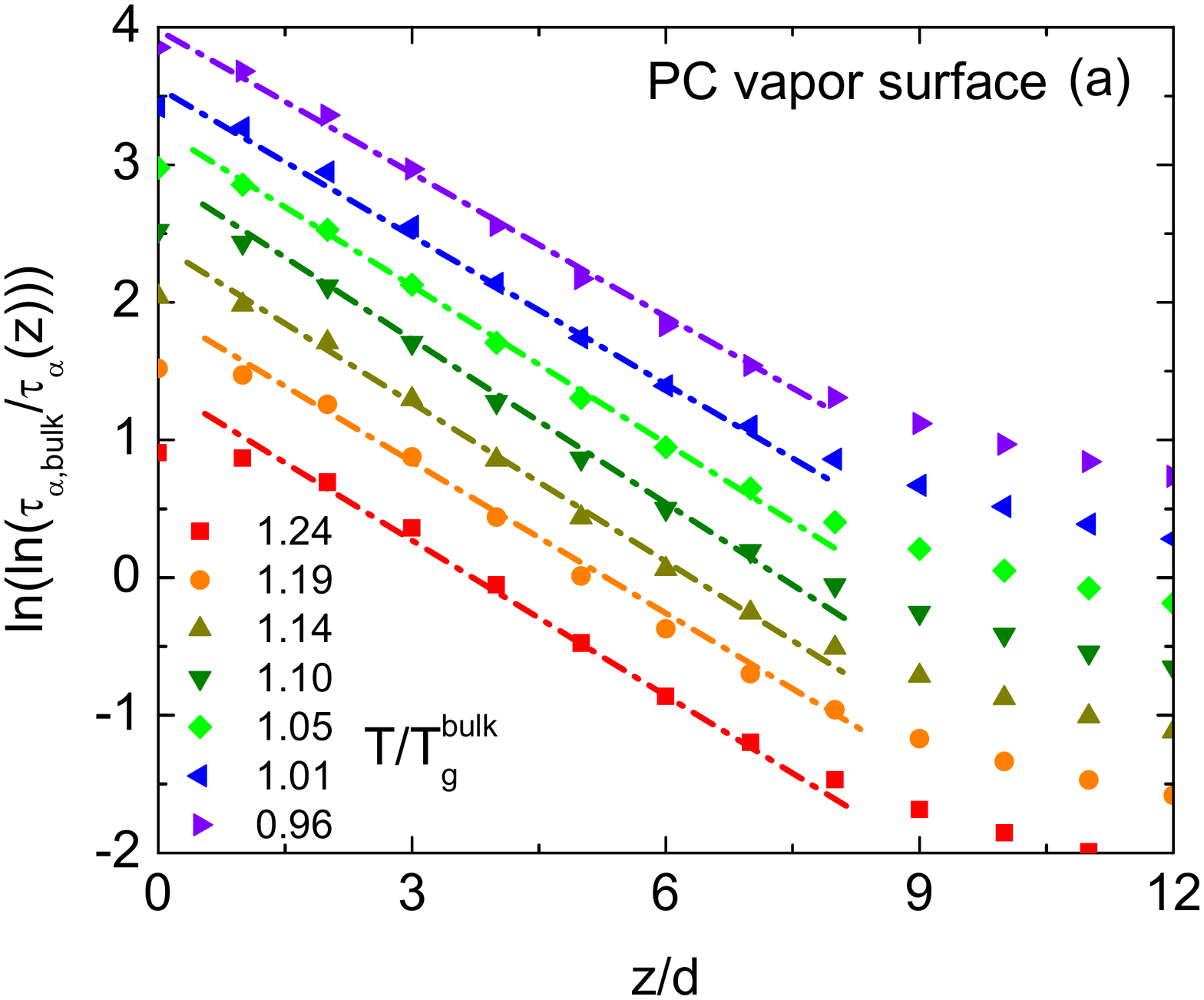}
\includegraphics[width=8.5 cm]{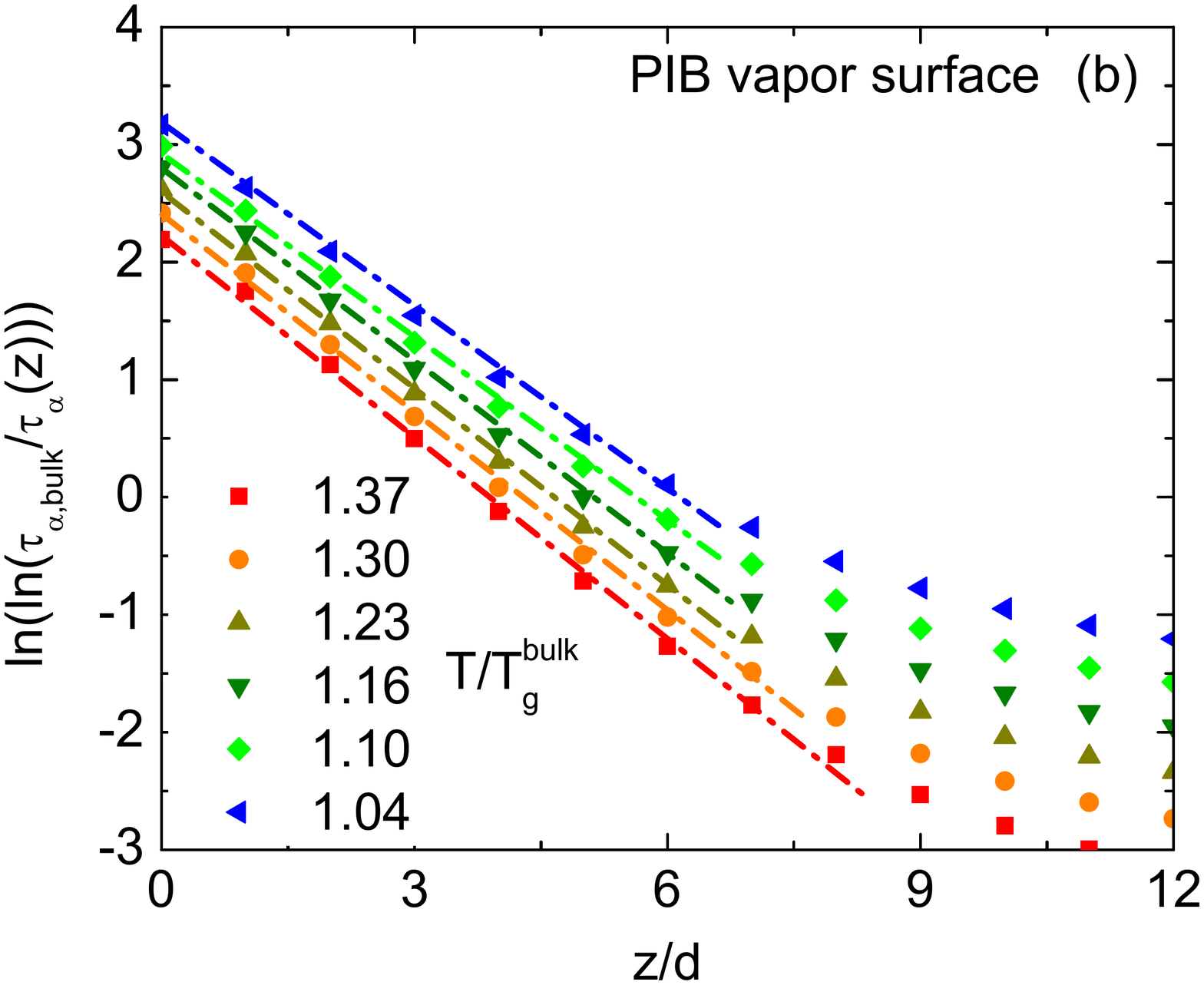}
\caption{\label{fig:5}(Color online) Double natural logarithm plot of the normalized inverse alpha relaxation time as a function of distance from the vapor surface over a wide range of reduced temperatures for (a) PC ($\xi$ increases from $3.1d$ to $3.8d$ with cooling) and (b) PIB polymer films ($\xi$ decreases from $2d$ to $1.7d$ with cooling). The dash-dot lines through the theoretical data points indicate a double exponential variation.}
\end{figure}

Figure 5 shows numerical results for polycarbonate (PC) and polyisobutylene (PIB) in the double-logarithmic vertical axis format over a wide range of temperatures in units of the bulk $T_g$. The differences between PC and PIB reflect their very different bulk dynamic fragilities of $m\sim 140$ and 46, respectively \cite{24,43,44}, and thus within ECNLE theory the \textit{relative} importance of the collective elastic versus local cage barrier. For low fragility PIB, the elastic barrier is far less important than the local cage barrier $F_B$ \cite{24,43}. The major trends in Figure 5 are the larger absolute enhancement, and more temperature sensitivity, of the alpha time at the vapor surface at fixed $T/T_g$ for a higher fragility system. Very good double exponential behavior for both systems extends to $\sim 8d$ ($d\sim$ nm is the Kuhn segment diameter \cite{23}) from the surface, and the decay lengths are all similar and vary very weakly with temperature. The bending up of the curves at large distances from the surface reflects the asymptotic approach to bulk behavior which is influenced by the cutoff of the elastic barrier at the vapor surface \cite{44}.
\subsection{Amplitude and Length Scales of Alpha Time Gradient}
The double exponential like behavior in Figures 4b and 5 can be quantitatively analyzed in two ways: fit our numerical calculations near interface ($z\leq 8d$) using eq 12 (per the main frame of Figure 4a) or fit to the double logarithmic form (per the inset of Figure 4a):
\begin{eqnarray}
\ln\left(\ln\left(\frac{\tau_{\alpha,bulk}}{\tau_{\alpha(z)}}\right)\right)=\ln A(\Phi)-\frac{z}{\xi(\Phi)}
\label{eq:13}
\end{eqnarray}
Since the theoretical (and also simulation) results are not of a perfect double exponential form, the extracted amplitude and length scale depend to some extent on precise fitting procedure.

For the hard sphere fluid, we find (not plotted) that the decay length varies by less than $10\%$ over the wide range of packing fractions studied. This variation is likely not significant given the double exponential form of the alpha time gradient is not exact, and hence modest ambiguity in the extracted value of the decay length is unavoidable. This viewpoint is buttressed by the fact we find the extracted $\xi(\Phi)$ weakly grows (decreases) from $\sim (2.3-2.6)d$ ($(3.2-2.8)d$) with increasing packing fraction of 0.55-0.62 based on the single (double) logarithmic fitting representation; overall, we deduce $\xi\sim 3d$.

Figure 6 presents, in three different formats, hard sphere fluid results for the amplitude which is essentially the difference between the total barrier in the bulk and at the vapor surface. Interestingly, the upper inset shows this amplitude grows roughly exponentially with packing fraction, $A(\Phi)=e^{-20.9}e^{39.54\Phi}$. Since NLE theory for the (transient) dynamic localization length in the bulk varies exponentially as \cite{53} $r_{L,bulk}/d=30e^{-12.5\Phi}$, a possible connection between $A$ and the dynamic localization length is \textit{empirically} suggested. This motivates the main frame plot in Figure 6 of $A$ versus the difference in the inverse dynamic localization length squared at the interface and in the bulk. We find a rough linear correlation, which is qualitatively in the spirit of phenomenological elastic models that posit the total barrier varies inversely with the square of the localization length \cite{19,47}. We are not advocating this connection is of fundamental importance. Rather, most significantly is the lower left inset of Figure 6 which shows that $A$ varies logarithmically with the bulk alpha time to a good approximation over 15 decades:
\begin{eqnarray}
A=a+b\log\left(\frac{\tau_{alpha,bulk}}{\tau_0}\right)
\label{eq:14}
\end{eqnarray}
where $a$ and $b$ are numerical constants. This result connects the surface-nucleated mobility enhancement amplitude with the bulk alpha time, a nontrivial link between bulk and interfacial dynamics. In section VI that addresses spatially inhomogeneous decoupling, the physical origin of eq 14 is discussed in detail. Briefly, we note that if the barrier factorization property of eq 11 applies, then eq 14 follows as:      $A=\log\left(\tau_{\alpha,bulk}/\tau_{\alpha}(z=0)\right)\propto\beta F_{total}^{bulk}-\beta F_{total}(z=0)\propto\beta F_{total}^{bulk}\left(1-f(z=0)\right)\propto\log\left(\tau_{\alpha,bulk}/\tau_{0}\right).$

\begin{figure}[htp]
\includegraphics[width=8.5 cm]{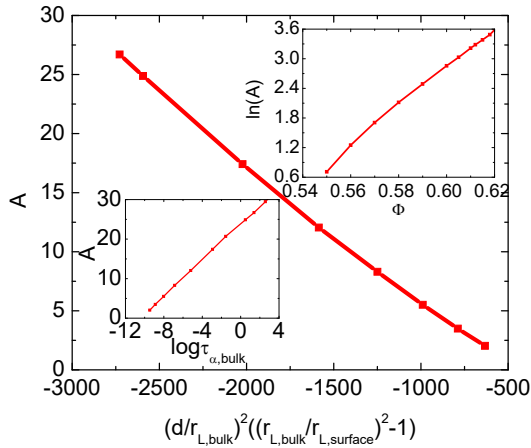}
\caption{\label{fig:6}(Color online) Mainframe: Amplitude of the double exponential relaxation time gradient of eq 12 as a function of the difference in the inverse square of the dimensionless bulk and surface localization lengths. Upper inset: natural logarithm of the amplitude versus packing fraction. Lower inset: Log-log (base 10) plot of the amplitude versus bulk alpha time.}
\end{figure}

Figure 7 presents the analogous results of the double exponential analysis of the alpha time gradient for the polymer liquids. The amplitude $A$  is quite insensitive to fitting procedure, and grows in a strongly nonlinear manner with inverse temperature for the fragile PC and PS systems, but in a nearly linear manner for low fragility PIB. In contrast, the decay length scale depends more on fitting procedure. However, for fixed chemistry, its overall variation with temperature is perturbative, with a material-specific magnitude that grows with fragility, varying from $\sim 1.8d$ for PIB to $\sim (2.6-3.6)d$ for PC. If one restores absolute units based on the mapped value of $"d"$ (known Kuhn segment diameters \cite{23,24,43,44}), these numbers are rather strongly nonuniversal, corresponding to $\sim 1.6$ nm for PIB, $\sim 3$ nm for PS, $\sim 4-5$ nm for PC

\begin{figure}[htp]
\includegraphics[width=8.5 cm]{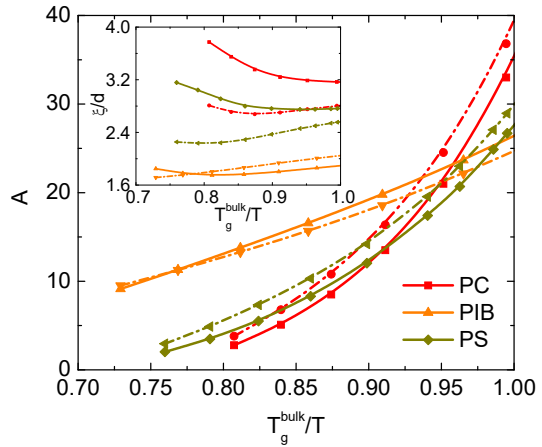}
\caption{\label{fig:7}(Color online) Amplitude (main frame) and decay length (inset) parameters in eq 12 as a function of $T_g^{bulk}/T$ extracted  based on fitting the theory numerical data in the format $\ln\left(\tau_{\alpha,bulk}/\tau_{\alpha}(z)\right)$ (solid curves) and $\ln\left(\ln\left(\tau_{\alpha,bulk}/\tau_{\alpha}(z)\right)\right)$ (dash-dot curves), respectively.}
\end{figure}

Overall, the theory predicts little temperature or density variation of the decay length. Simulations in the lightly supercooled regime \cite{13,39,40,42} usually find a modest growth with cooling, to varying \textit{model-dependent} degrees. For example, the ZM \cite{37} and H \cite{51} polymer simulations found $\xi$ grows by a factor of less than 2 over the range studied, reaching $\xi\approx 2d$ at the lowest temperature. The RR simulations \cite{52} based on a different model and fitting procedure to extract a length scale found $\xi\approx 4.3d$. Most significantly, Simmons et al \cite{41} have recently shown, for the first time, such a length ($\sim 3-5$ in their units) \textit{saturates} with cooling in thick free-standing films. This new work agrees with prior simulations performed over a more limited range of time scales where the length scale does grow weakly. But, very importantly, ref \cite{41} goes to significantly longer time scales than prior work thereby revealing the saturation behavior and seemingly establishing the limiting low temperature behavior of the penetration length scale in films with a vapor interface.

\section{Alternative Growing Length Scales and Interfacial Layer-Averaged Time Scale}
Given present experimental limitations render high resolution extraction of the alpha time gradient apparently impossible, a direct experimental test of our predicted double exponential form and the associated variations of amplitude and length scale does not seem possible. Hence, it is important to ask what are the distinctive consequences of this form of the spatial gradient for observable properties that average over it in various ways. This is focus of sub-sections IVB and IVC below. But we first consider the question of a characteristic practical length scale of the alpha time gradient which can be easily deduced from computer simulations.
\subsection{Perturbed Layer Size via Bulk Relaxation Time Recovery Criterion}

Given there are spatial gradients of all quantities in a film, as a matter of principle no unique dynamical length scale can be defined. Based on the double exponential form of the alpha time gradient, in our theory the characteristic length scale, $\xi$, is only very weakly temperature or density dependent. This reflects the basic theoretical ideas as formulated at the dynamic free energy level \cite{25}. However, the caging component of the dynamic free energy is, of course, not directly measureable, but rather only its dynamical consequences. Even at the level of the alpha time gradient, one can define a dynamic length scale in different ways. A common one is to define a "perturbed layer thickness", $\zeta$, based on a recovery of bulk relaxation time criterion $\tau_\alpha(z=\zeta)/\tau_{alpha,bulk}=C$ \cite{25,35,37,49}. This length scale should not be confused with a “mobile layer thickness” defined as the spatial region near a surface that remains “liquid-like”. The latter length grows with heating, in contrast with $\zeta$ which grows with cooling. 

Representative results for $\zeta$  are shown in Figure 8 for PS and $C= 0.5$, with and without the new mobility transfer physics. Comparison with the simulations of Simmons and coworkers \cite{41} are also shown, and seem to agree well with the theory in the limited range probed. For low temperatures or long relaxation times, the large length scales predicted reflect the presence of the subtle long range tail of the mobility gradient due to the cutoff at the surface of the collective elastic field contribution to the total barrier \cite{25,44,49,50}. The more local mobility transfer effects strongly enhance the length scale at high temperatures, but the overall shape of this mobile layer variation is largely unchanged from prior work. 

\begin{figure}[htp]
\includegraphics[width=8.5 cm]{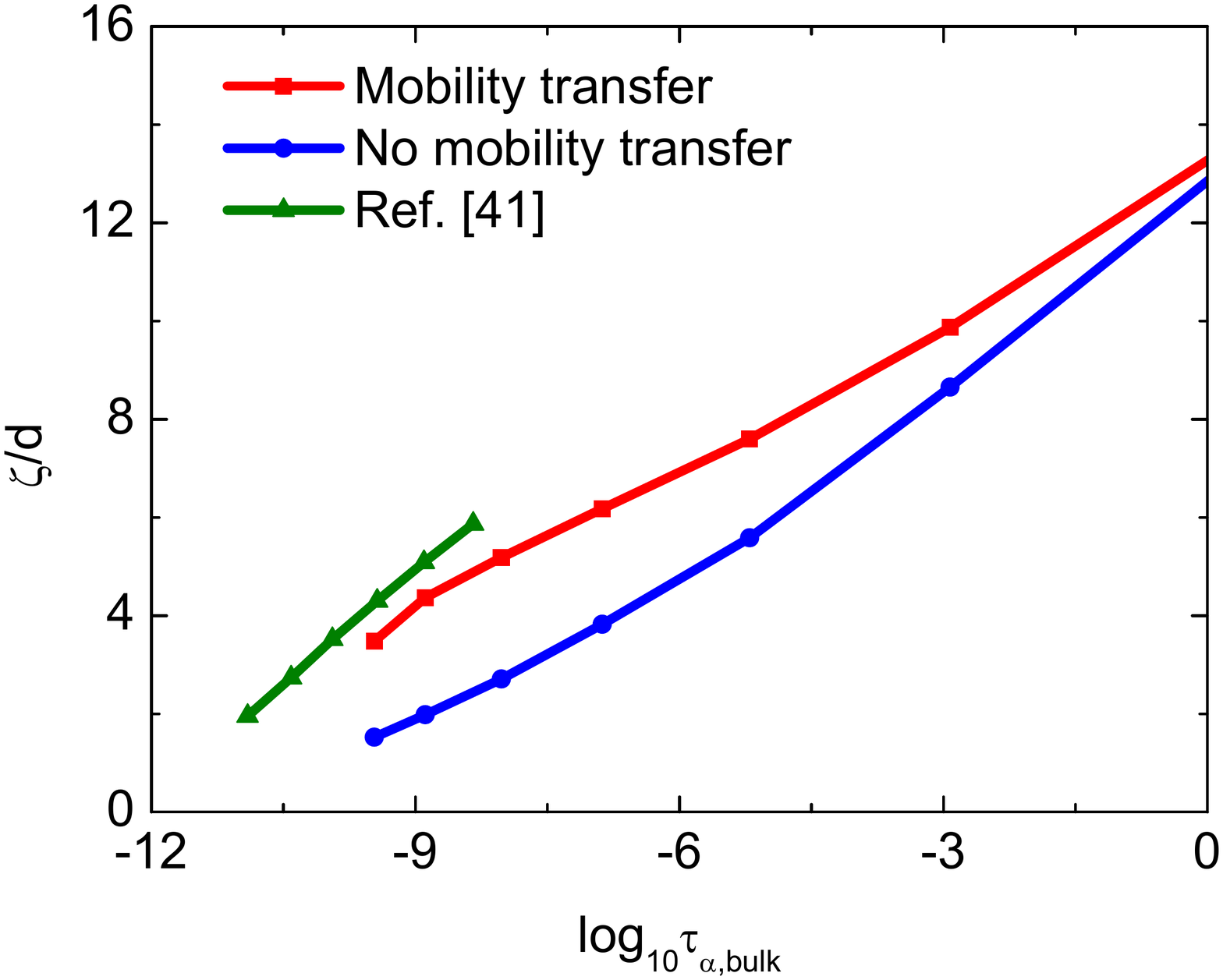}
\caption{\label{fig:8}(Color online) Dimensionless perturbed layer thickness computed based on the criterion $\tau_\alpha(z=\zeta)/\tau_{\alpha,bulk}=0.5$ as a function of $\log_{10}\tau_{\alpha,bulk}$ for PS melt with and without mobility transfer effects. Also shown are simulation results41 extracted using our $C=0.5$ criterion.}
\end{figure}

\subsection{Practical Interfacial Layer Thickness}
The perturbed layer length discussed in the previous section is generally not directly measurable in experiment. We thus explore an alternative "mean interfacial length", $L_{int}$, motivated by recent dielectric spectroscopy measurements that extract it \cite{54,55}. We use the criterion $\tau_\alpha(z=L_{int})/\tau_{\alpha,bulk}=C$ to define this length scale, but consider the practical experimental sensitivity situation where the long range, but very low amplitude, tail of the mobility gradient due to elastic field cutoff effect is not probed. To allow analytic insights that incur little quantitative error, we adopt $C=1/e$ and use eq 13 with $z=L_{int}$ which defines an interfacial length as:
\begin{eqnarray}
L_{int}=\xi(\Phi)\ln{A(\Phi)}
\label{eq:15}
\end{eqnarray}
Since $\xi(\Phi)$ is weakly temperature or packing fraction dependent, we set $\xi(\Phi)\approx 3d$. We emphasize that the behavior of $L_{int}$ is \textit{directly} related to the double exponential form of the alpha time gradient.

Calculations of the temperature dependence of $L_{int}$ for PS, PIB and PC are shown in Figure 9. Note that $L_{int}(T)$ grows significantly with cooling, and in a roughly linear manner for PS and PIB. Interestingly, the same qualitative behavior has been experimentally observed \cite{54,55} for polymers near solid surfaces using dielectric loss spectroscopy. We suggest these predictions can be tested experimentally for a vapor interface system using dielectric spectroscopy or other methods (perhaps per ref \cite{26}). The origin of the theoretical behavior is the strong temperature dependence of the amplitude, $A$, of the normalized alpha time gradient, not the intrinsic length scale. Its roughly linear temperature variation appears to be understandable as follows. From our previous work \cite{43}, the mapping from thermal liquids to hard spheres corresponds to the connection $T(\Phi)=T_{reff}-(\Phi-0.5)/\alpha_{T}$, where $T_{reff}$ depends on the material and $\alpha_T$ is the liquid thermal expansion coefficient. Given this, and our finding above concerning $A$, one can write $A(\Phi)=e^{-20.9}e^{39.54\Phi}=e^{-20.9}e^{39.54\left[(T_{reff}-T)\alpha_T+0.5\right]}=e^{-1.13}e^{39.54(T_{reff}-T)\alpha_T}$ , or $\ln A(\Phi)=-1.13+39.54\alpha_T\left(T_{reff}-T\right)$. This result plus eq 15 qualitatively explains the origin of the near linear growth of $L_{int}(T)$  with temperature found numerically. 

\begin{figure}[htp]
\includegraphics[width=8.5 cm]{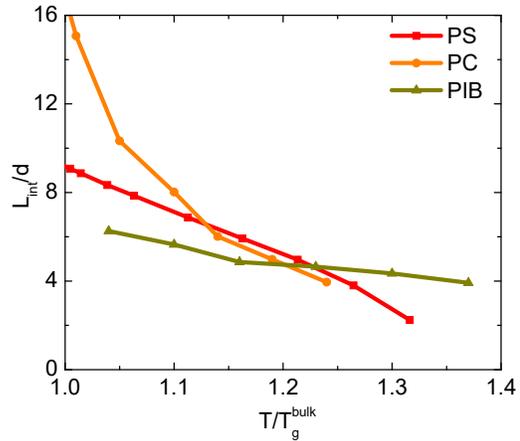}
\caption{\label{fig:9}(Color online) The reduced interfacial layer thickness, $L_{int}(T)$, of PS, PC, and PIB thick films in units of the Kuhn segment diameter as a function of $T/T_{g}^{bulk}$. The curves through the discrete theoretical data points are a guide-to-the-eye.}
\end{figure}

\subsection{Practical Layer-Averaged Time Scale}
Experiments usually only measure ensemble-averaged correlations functions in the time or frequency domain and thus perform a spatial average over the relaxation time spatial gradient. We analytically consider this problem based on a generic model \cite{49,50} that we believe is relevant to measurements such as dye-reorientation-induced decay of the fluorescence intensity \cite{26} and dielectric spectroscopy. Here we analyze the problem in the frequency domain.

We suppose the relevant time correlation function, $C(t)$, is a simple arithmetic average over the exponential relaxation contributions of molecules or segments in the gradient region of width defined by the practical interfacial layer size discussed above:
\begin{eqnarray}
C(t)=\frac{1}{L_{int}}\int_{0}^{L_{int}}dz\left<e^{-t/\tau_{alpha}(z)}\right>
\label{eq:16}
\end{eqnarray}
To make contact with frequency domain measurements such as dielectric spectroscopy, we calculate the corresponding loss function:
\begin{eqnarray}
C''(\omega)=\frac{1}{L_{int}}\int_{0}^{L_{int}}dz\frac{\omega\tau_\alpha(z)}{1+\left(\omega\tau_\alpha(z)\right)^2}
\label{eq:17}
\end{eqnarray}
Non-dimensionalizing frequency and relaxation time as $\Omega=\omega\tau_{\alpha,bulk}$ and $B(z)=\frac{\tau_\alpha(z)}{\tau_{\alpha,bulk}}=e^{-A(\Phi)e^{-z/\xi(\Phi)}}=e^{e^{-L_{int}/\xi}e^{-z/\xi}}$, respectively, one can write:
\begin{eqnarray}
C''(\omega)=C''(\Omega)=\frac{1}{L_{int}}\int_{0}^{L_{int}}dz\frac{\Omega B(z)}{1+\left(\Omega B(z)\right)^2}
\label{eq:18}
\end{eqnarray}
We then define the layer-averaged interfacial relaxation time as the inverse of the maximum frequency of the loss function, $\left<\tau_{int}=1/\omega_{max}\right>$ , which is determined by the equation: 
\begin{eqnarray}
\frac{\partial C''(\Omega)}{\partial\Omega}|_{\Omega_{max}}&=&\frac{1}{L_{int}}\int_{0}^{L_{int}}dzB(z)\nonumber\\
&\times&\frac{1-\left(\Omega_{max}B(z)\right)^2}{\left[1+\left(\Omega_{max}B(z)\right)^2\right]^2}=0
\label{eq:19}
\end{eqnarray}
Changing integration variables to $u=B(z)$, one obtains a closed nonlinear equation for $\Omega_{max}=\omega_{max}\tau_{\alpha,bulk}=\tau_{\alpha,bulk}/\left<\tau_{int}\right>$:
\begin{eqnarray}
0&=&\frac{\xi}{L_{int}}\int_{1/e}^{e^{-e^{L_{int}/\xi}}}\frac{du}{\ln u}\frac{1-\left(\Omega_{max}u\right)^2}{\left[1+\left(\Omega_{max}u\right)^2\right]^2}\nonumber\\
&\approx&\int_{1/e}^{0}\frac{du}{\ln u}\frac{1-\left(\Omega_{max}u\right)^2}{\left[1+\left(\Omega_{max}u\right)^2\right]^2}\
\label{eq:20}
\end{eqnarray}
The second form follows since $L_{int}/\xi(\Phi)=\ln A(\Phi)$. For $\Phi\sim 0.58-0.62$, $\ln A(\Phi)$  increases from $\sim 2.2$ to 3.6, implying $e^{-e^{L_{int}/\xi(\Phi)}}\approx 0$ to a good approximation. Equation 20 corresponds to a \textit{universal} formula for the layer-averaged alpha time in the deeply supercooled regime.  Solving it numerically yields: 
\begin{eqnarray}
\frac{\left<\tau_{int}\right>}{\tau_{\alpha,bulk}}\approx 0.12
\label{eq:21}
\end{eqnarray}

The above analysis leads to the striking prediction that, to leading order, the \textit{layer-averaged} relaxation time is only modestly reduced by a factor of $\sim 8$ \textit{relative} to its bulk value, and in a system and temperature or density \textit{independent }manner, despite the presence of enormous and chemically-specific local relaxation time gradients close to the surface. Such behavior might seem counterintuitive. It is partially a consequence of the ensemble averaging in the time \cite{56} or frequency domain being generically weighted towards the slower relaxing regions of the film. But, even more crucially, our above mathematical analysis shows this result is intimately connected with the double exponential form of the alpha time gradient. Our result also serves as warning - observing a nearly temperature invariant and modest \textit{average }enhancement of the interfacial relaxation time relative to its bulk analog does not imply there are no massive changes of the alpha time near an interface. Recent experiments by Sokolov et al \cite{54,55} on nanocomposites where glass-forming liquids are in contact with large spherical particles appear consistent with our analysis in that the \textit{ratio} of the interfacial alpha time to its bulk analog is found to be nearly temperature-independent and only modestly larger than unity. Even though we have not analyzed the problem of glass-forming liquids near a solid interface in the present article, the striking result of Eq 21 largely transcends the precise nature of the interface and follows from the double exponential form of the alpha time gradient. In a future article, we will explicitly show that our theory does indeed predict a double exponential relaxation time gradient for thick films with one solid (rough or smooth) surface. 
\section{Spatial Variation of the Alpha Time Temperature Dependence and Local $T_g$ Gradients}
An interesting fundamental question is how the temperature dependence of the alpha relaxation time varies with distance $z$ from the vapor interface. Figure 10 shows representative calculations that address this for PS. The rate of increase of the relaxation time with cooling is very strongly suppressed near the interface, corresponding to a reduction of dynamic fragility upon moving towards the vapor surface \textit{if} fragility is evaluated at the \textit{bulk} $T_g$. If a $z$-dependent fragility is determined based on $T_g(z)$ and not the bulk $T_g$, then far less change relative to its bulk value is expected. Bulk-like temperature dependence behavior is recovered at roughly $\sim 12$ nm from the surface for PS.

Motivated by critical issues in the formation of vapor-deposited ultrastable glasses \cite{57}, Figure 11 presents the alpha time calculations of Figure 10 in the top 3 layers extended to temperatures far below the bulk $T_g$, assuming the system remains in equilibrium for those value of $z$. We find that these layers remain liquid down to $\sim 0.79$ and 0.84 of the bulk glass transition temperature for the top and second layer, respectively. The essential physics is that the onset of strong caging, high barriers and activated motion is shifted to much lower temperatures due to the combined consequences of three effects: (i) loss of neighbors at the vapor interface which at a fixed temperature massively weakens the localized form of the dynamic free energy relative to bulk behavior, (ii) cutoff of the elastic barrier is maximized very close to the vapor interface, and (iii) the longer range mobility transfer effect further reduces the local cage and elastic barriers, with the latter occurring since reduction of the jump distance and elastic modulus softening penetrates deeper into the film. But, at sufficiently low temperature (well below the bulk $T_g$), the dynamic caging constraints inevitably become strong if the system is equilibrated, consistent with studies of ultra-stable glass formation \cite{57}. Thus, strongly temperature-dependent high barriers must eventually emerge, implying relaxation near the surface becomes strongly non-Arrhenius. 

The dashed horizontal lines in Figure 11 indicate time scales typical of the deposition rate employed in the vapor deposition of ultrastable films \cite{57}. Our prediction that the top two layers remain equilibrated (defined as an alpha time less than 100 s) on the typical deposition time scale seems consistent with the experimental observation \cite{57} that the maximum densification of ultrastable films is achieved when the substrate temperature is of order $80\%$ of the bulk $T_g$. We do caution that, beyond the obvious fact the theory is approximate, our results should likely not be over-interpreted quantitatively given our adoption of a simple molecular model, a sharp vapor-liquid interface model, and other simplifications. The results in Figure 11 are also relevant for the massively accelerated “surface diffusion” phenomenon seen experimentally \cite{57}, but this topic is beyond the scope of the present article.

\begin{figure}[htp]
\includegraphics[width=8.5 cm]{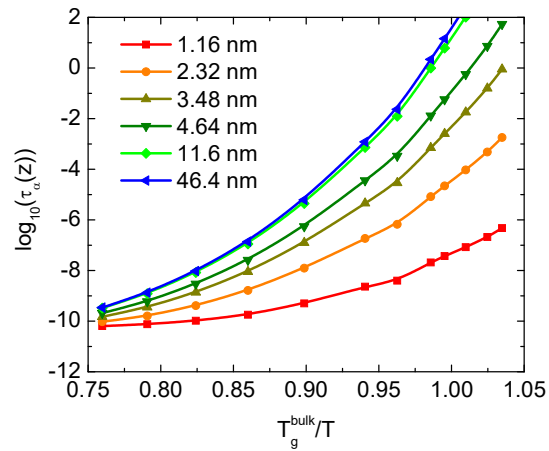}
\caption{\label{fig:10}(Color online) Angell-like plot of the alpha time (in seconds) at various indicated distances $z$ from the vapor surface of a PS ﬁlm. The effective hard sphere (Kuhn segment) diameter for PS is $d\sim 1.16$ nm.}
\end{figure}

\begin{figure}[htp]
\includegraphics[width=8.5 cm]{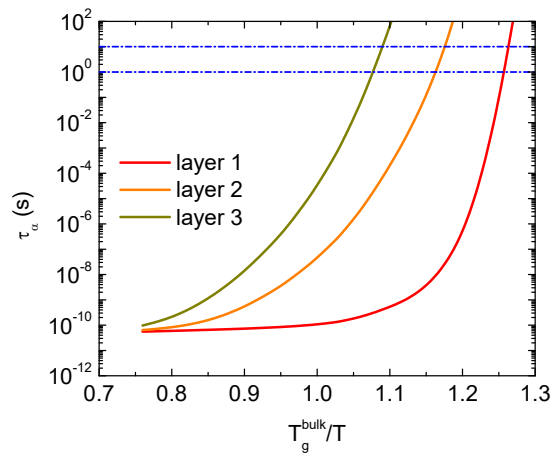}
\caption{\label{fig:11}(Color online) Extension of the calculations of Figure 10 to far below the bulk glass transition temperature for the alpha relaxation time in the top 3 layers of the film. The horizontal blue dashed lines indicate time scales relevant to typical deposition times in the fabrication of ultra-stable films \cite{57}.}
\end{figure}

Based on calculations like those in Figures 10 and 11, one can compute a $z$-dependent glass transition temperature. Figure 12 presents results for the normalized $T_g$ gradient, and also that of the prior ECNLE theory without the mobility transfer physics \cite{44}. Calculations are shown for the common experimental vitrification criterion of an alpha time of 100 s, and also, to connect with simulations, the corresponding predictions based on a short 100 nsec criterion. Results with and without the new surface-mediated mobility transfer effect are also shown. We note that experimental methods do presently exist and are being further developed \cite{2,5,33} to extract the glass transition temperature gradient.

There are several notable features in Figure 12. (i) Regardless of which version of ECNLE theory is adopted, the \textit{normalized} $T_g(z)/T_g^{bulk}$ gradient is only modestly sensitive to the vitrification time scale criterion. The physical reason is discussed below. (ii) As expected, the addition of the longer range mobility transfer physics renders the gradient much more slowly decaying in space, with $T_g$ reductions extending $\sim 12$ nm ($\sim 10d$) into the bulk. It also qualitatively changes visually the "2-regime" shape predicted by the prior ECNLE theory \cite{44,49,50}, with the spatial variation now much smoother. (iii) Theory and the PS atomistic simulations of Zhou and Milner \cite{37} agree well with the theory for the large $T_g$ suppression very close to the surface, although bulk behavior is recovered on a shorter length scale ($\sim 5$ nm) in simulation than predicted. Whether the latter reflects differences in the models analyzed, time scales accessible (nsec vs 100 s), and/or errors incurred by the model simplifications and theoretical approximations, requires further study. (iv) Empirically, we find our calculations are well fit by: 
\begin{eqnarray}
\frac{T_g(z)}{T_g^{bulk}}&=&1-\frac{0.859}{z/d+56.56}-\frac{0.0222}{\left(z/d+0.7026\right)^2}\nonumber\\
&-&0.28035e^{-0.5088z/d}
\label{eq:22}
\end{eqnarray}

\begin{figure}[htp]
\includegraphics[width=8.5 cm]{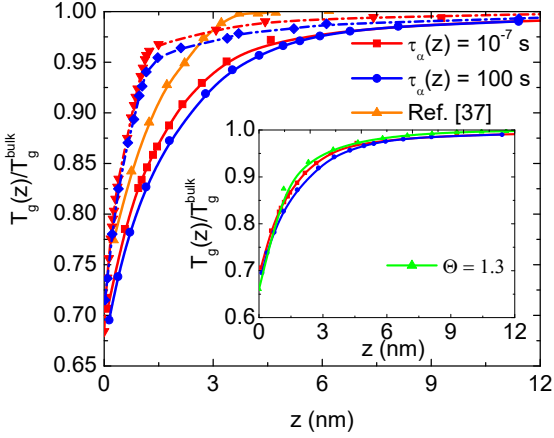}
\caption{\label{fig:6}(Color online) Local $T_g$ in a PS film divided by its bulk value as a function of distance from the vapor interface as computed based on the indicated two different vitrification criteria. Also shown (orange triangles) are simulation data \cite{37} based on a vitrification time scale criterion of 100 ns. The solid and dashed-dotted curves correspond to theory calculations with and without mobility transfer, respectively. The empirical fit function for the theoretical normalized $T_g$ gradient is given by eq 22. Inset: normalized $T_g$ gradients calculated with mobility transfer effects (red and blue points and curves per the mainframe) and the analytical expression of eq 29 and 30 (green points and curve) with the indicated value of $\Theta$ and $f(z)=1-0.904e^{-z/2.644d}$ as discussed in the caption of Figure 3 for $\Phi=0.61$.}
\end{figure}

Analytic insight can be obtained concerning the functional form and physical origin of the normalized $T_g$ gradient, and how it depends on the vitrification time scale criterion. The analysis below applies, to leading order, to \textit{any} interface. A caveat is that it is not "exact" within ECNLE theory, but is reliable based on our numerical studies. First recall we numerically found that the theory predicts the total barrier in films \textit{factorizes} to a good approximation into a product of its bulk liquid value multiplied by a temperature and chemistry \textit{independent} function of location in the film:
\begin{eqnarray}
F_{total}^{film}(T,z)\approx f(z)F_{total}^{bulk}(T)
\label{eq:23}
\end{eqnarray}
where $f(z)\rightarrow 1$ on a scale o $f\sim 3d$ in an exponential manner (Figure 3). Recall this quasi-universal factorization property and the near exponential variation of $f(z)$ is what underlies our prediction of \textit{both} a double exponential form of the alpha time gradient and the “decoupling” effect discussed in the next section. Now, the glass transition at a location $z$ from the surface or in the bulk can be defined by a variable time scale criterion set by a number $"y"$ as
\begin{eqnarray}
\tau_{\alpha,film}\left(T_g(z)\right)=10^ys=\tau_{\alpha,bulk}\left(T_{g,bulk}\right)
\label{eq:24}
\end{eqnarray}\\
To leading order, this criterion is equivalent to a constraint on the total barrier, $F_{total}^{film}\left(T_g(z)\right)=F_{total}^{bulk}\left(T_{g,bulk}\right)=b$ , which implies:
\begin{eqnarray}
\frac{F_{total}^{bulk}\left(T_{g,bulk}\right)}{F_{total}^{film}\left(T_g(z)\right)}=f(z)	
\label{eq:25}
\end{eqnarray}
where $b$ is in units of the thermal energy and can be as low as $~\sim 10$ in simulations and $\sim 32$ for experiments corresponding to $\sim 4-5$ or 14 decades variation of the relaxation time, respectively. 

The factorization property of eq 25 implies one \textit{only} needs to know the temperature dependence of the total dynamic barrier in the \textit{bulk} liquid, which has been addressed in prior ECNLE theory work \cite{19,20}. For fragile liquids and alpha times spanning the enormous range of $\sim 1$ ns to 100 s, the so-called "parabolic law" \cite{58,59} empirically captures the numerical predictions of ECNLE theory \cite{20,21} for the alpha time:
\begin{eqnarray}
\log\left(\frac{\tau_\alpha(T)}{\tau_0}\right)=\left(\frac{J}{k_BT_0}\right)^2\left(1-\frac{T_0}{T}\right)^2,\quad T\leq T_0	
\label{eq:26}
\end{eqnarray}
where $J$ and $T_0$ are a system-specific energy and “onset” or crossover temperature, respectively. An explicit formula for the short time scale, $\tau_0$, can be written down \cite{19,20,21}, and is a very weak function of temperature. For our present approximate analytic analysis purposes, $\tau_0$ is not needed. Importantly, eq 26  (with the assumption of a constant $\tau_0$) has also been shown to fit well large quantities of experimental data of diverse glass-forming liquids in the deeply supercooled regime \cite{58,59}. Hence, the form of eq 26 applies in a practical sense independent of ECNLE theory. We then define an effective total barrier as $\ln\left(\frac{\tau_\alpha(T)}{\tau_0}\right)\approx\beta F_{total}(T)$. Then, using eqs 25 and 26, and recalling that we showed the total normalized barrier is to zeroth order equal to $f(z)$, one obtains:
\begin{eqnarray}
f(z)=\left[\frac{1-\frac{T_0}{T_{g,bulk}}}{1-\frac{T_0}{T_g(z)}}\right]^2	
\label{eq:27}
\end{eqnarray}
The energy scale $J$ cancels out. Defining a normalized $T_g$ gradient variable, $Y=T_g(z)/T_{g,bulk}$, yields: 
\begin{eqnarray}
f(z)=\left[\frac{1-\Theta}{1-\frac{\Theta}{Y(z)}}\right]^2,\quad\Theta-\frac{T_0}{T_{g,bulk}}>1
\label{eq:28}
\end{eqnarray}
from which an analytic relation that connects the normalized $T_g$-gradient and $z$-dependence of the total barrier is obtained: 
\begin{eqnarray}
Y(z)=\frac{\Theta}{1+\frac{\Theta-1}{\sqrt{f(z)}}}
\label{eq:29}
\end{eqnarray}
Obviously, if $f(z)\rightarrow 1$ (as it must at large $z$), then $Y\rightarrow 1$.

Equation 29 is an almost universal relation except for the ratio, $\Theta$, of the onset to glass transition temperatures. It can depend on chemistry (increases as fragility decreases), pressure, \textit{and} the vitrification time scale criterion \cite{24,43}. But we recall that ECNLE theory predicts $T_{g,bulk}$ varies roughly logarithmically with time scale criterion \cite{43}, so $\Theta$ is expected to be weakly dependent on vitrification criterion (consistent with normalized $T_g$ gradients being weakly dependent on vitrification criterion). The inset of Figure 12 shows that eq 29 rather accurately describes our numerical results based on a sensible value \cite{20,57} of $\Theta$, thereby providing the sought after analytic insight concerning the functional form and physical origin of the $T_g$ gradient. The accuracy of eq 29 in reproducing our full numerical results also further supports the key result of eq 25.

The above analysis appears to explain, at zeroth order, why we numerically predict that the time scale criterion has a very minor effect on the \textit{normalized} $T_g$ gradient. More generally, we suggest it explains the often rather good and very surprising agreement between simulations based on a short vitrification time scale criteria (e.g., $\sim 1$ ns - 100 ns) and experimental measurements based on the 100 sec criterion for the \textit{normalized} $T_g$ gradient.  Finally, and very important for experimental tests of our ideas, we have shown there is a direct link between the functional form of the glass transition gradient and the theoretical concepts of total barrier factorization, double exponential form of the alpha relaxation time gradient, and the decoupling effect which we now discuss. 

\section{Spatially Inhomogeneous Dynamic Decoupling }
We now consider the question of spatial “decoupling” of the alpha time from its bulk value in thick films near a vapor interface, a striking phenomenon recently discovered using simulation by Simmons and coworkers \cite{41}. The latter workers performed simulations to significantly longer times than prior studies, which allowed them to discover that the alpha time gradient decay length and $z$-dependent decoupling exponent saturate at low enough temperature which is still far above the laboratory glass transition temperature defined by a 100 s criterion. Such a decoupling phenomenon might be theoretically expected based on our recent work \cite{25} that showed eq. 25 holds to a good approximation for the local cage barrier. However, in ECNLE theory there are also collective elastic effects. But given that the essential elements needed to quantify it (localization well curvature, jump distance) also follow from the local NLE dynamic free energy, effective factorization still seems plausible. 
	
Figure 13a presents in a double logarithmic format the time scale ratio $\tau_\alpha(z)/\tau_{\alpha,bulk}$ versus $\tau_{\alpha,bulk}$ for a wide range of distances from the interface over an exceptionally broad range of the bulk relaxation times (15 decades) from 0.1 nsec to $10^5$ s. One sees the theoretical data are very well described as straight lines, thereby indicating power law behavior characterized by a fractional exponent that depends on location in the film, $z$. To mathematically express this behavior requires non-dimensionalizing $\tau_{\alpha,bulk}$ by a temperature-independent constant time scale, $\tau_0^{*}$, the choice of which is not crucial for the $z$-dependent power law behavior one sees in Figure 13a. For convenience we choose it to be 1 s, and define the power law decoupling relation in a proportionality sense as:   
\begin{eqnarray}
\frac{\tau_\alpha(z,T)}{\tau_{\alpha,bulk}(T)}\propto\left(\frac{\tau_{\alpha,bulk}(T)}{\tau_0^{*}}\right) ^{-\varepsilon(z)}
\label{eq:30}
\end{eqnarray}
where $\varepsilon(z)$ is the $z$-dependent “decoupling exponent” function. If eq 30 was exact, it would essentially indicate an exact factorization of the dynamic barrier in films into a product of a $z$-dependent function times the bulk temperature-dependent barrier, and vice-versa. This behavior was previously predicted by the ECNLE theory without mobility transfer physics \cite{44}, indicating the conceptual foundation of such decoupling relates to the most general aspect of ECNLE theory of films, the idea of a position-dependent dynamic free energy \cite{49,50}. The new theoretical results in Figure 13a show decoupling is again predicted including the mobility transfer effect. Remarkably, to an excellent approximation a single $z$-dependent power law goes through all the data points spanning 15 decades.

\begin{figure}[htp]
\includegraphics[width=8.5 cm]{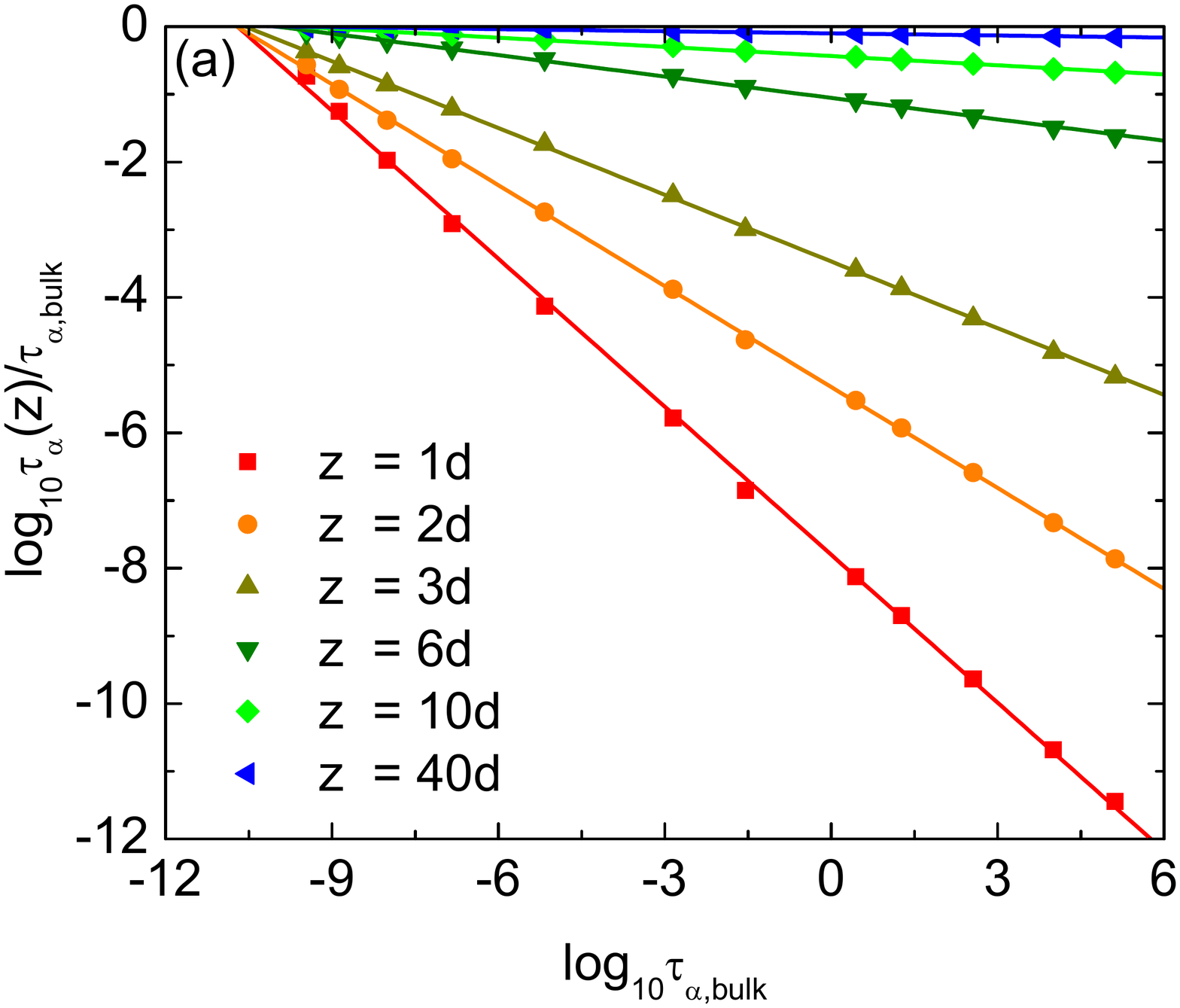}
\includegraphics[width=8.5 cm]{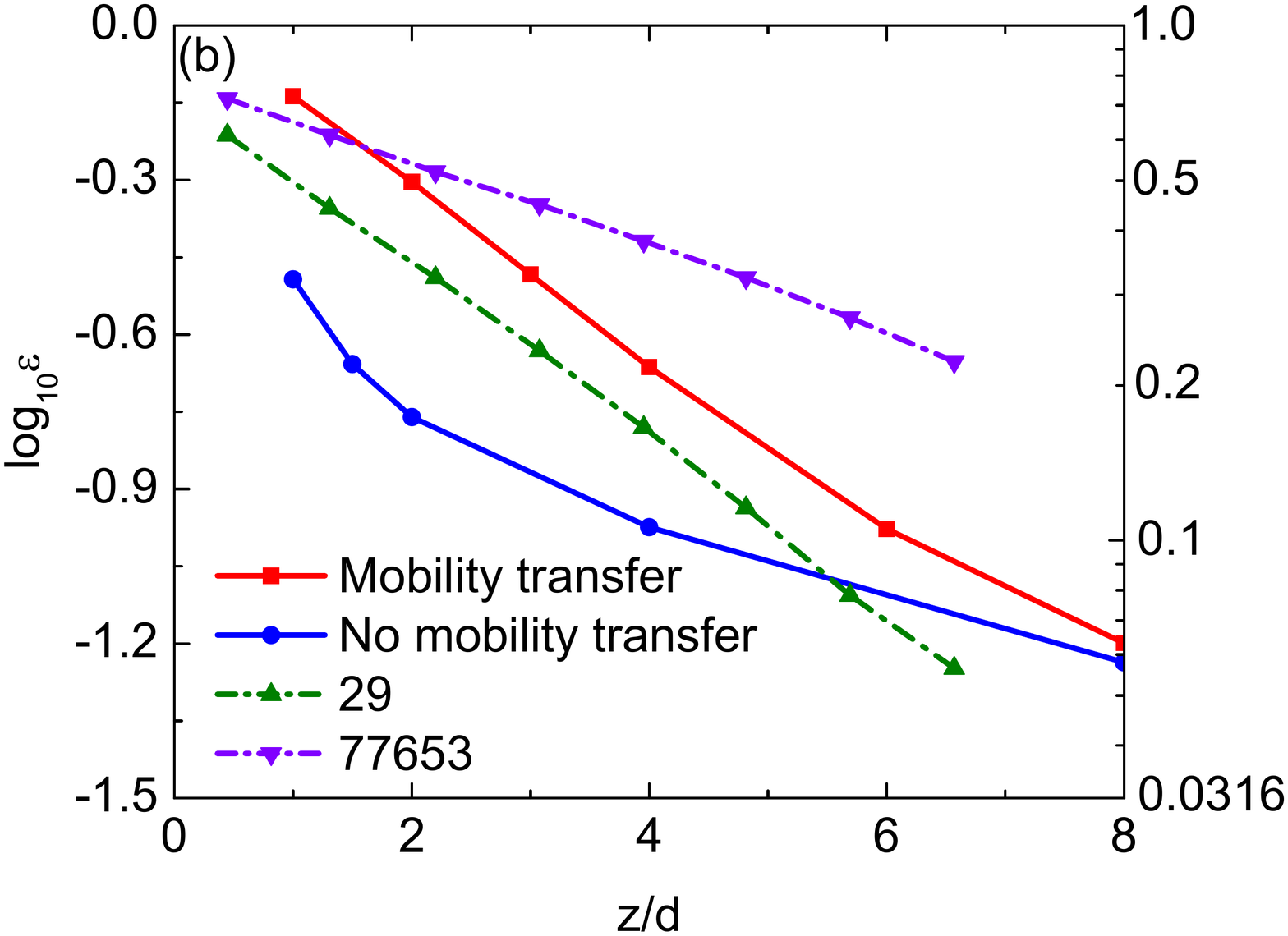}
\caption{\label{fig:13}(Color online) (a) Log-log (base 10) plot of $\tau_\alpha(z)/\tau_{\alpha,bulk}$ versus the bulk alpha relaxation time for a PS thick film at several distances from the vapor surface. The lines are power law ﬁts through all the theoretical data points which span 15 decades in bulk alpha time. (b) Decoupling exponent function (red) determined by fitting the theoretical data points which span the ﬁrst 5 decades of bulk alpha time (corresponds to “simulation time scale” \cite{41}), as a function of dimensionless inverse distance from the surface; analogous results not including the mobility transfer effect are shown in blue. The dash-dot curves are simulation data \cite{41} at several bulk relaxation times (indicated as 29 and 77653 expressed in a LJ model time unit). The theoretical results (red data points) are rather well fit over the $z$-range shown to an exponential function (not shown) given by $\varepsilon(z)=1.1e^{-0.42z/d}$.}
\end{figure}

The results in Figure 13a can be used to extract the decoupling exponent function. Figure 13b shows calculations based on the new theory and the prior ECNLE theory \cite{44} without the mobility transfer effect. Although it does not really matter given the results in Figure 13a, we extract the decoupling exponents using just the first 5 decades (covers the bulk relaxation time ranging from 0.1 ns to $10^4$ ns) to compare with recent simulation data \cite{41}. As expected, for all locations in the film shown in Figure 13b, the decoupling exponent with the mobility transfer physics is larger than its analog that ignores this effect. The difference is much larger near the surface, and monotonically extends far deeper into the film for the new theory. Moreover, the new theory predicts a nearly exponential variation of the decoupling exponent with location in the film given by $\varepsilon(z)=1.1e^{-0.42z/d}$ corresponding to a decay length of $\sim 2.4d$. This functional form is in qualitative contrast to the prior theory \cite{44} with no mobility transfer effects. The new theory is in good accord with simulation \cite{41} up to $z\sim 7d$, as shown by the curves based on relatively short and long dimensionless time scale criteria used to extract simulation alpha times. 

At very large $z$, the theory predicts the cutoff at the surface of the collective elastic field becomes important, and the decoupling exponent follows an inverse power law decay in $z$ form as previously found \cite{44} (this is why the red and blue curves tend to converge in Figure 13b). This crossover occurs at rather small values of $\varepsilon\sim 0.05-0.1$ corresponding to $z/d\sim 6-8$.

To avoid any confusion regarding our use of the word “decoupling”, and to provide further rationale for adopting this jargon, we note that a power law like formula per eq 30 is what is typically meant means by decoupling in bulk glass forming liquids \cite{10}. Specifically, the analog of the left hand side is the ratio of a characteristic translational time scale (deduced, e.g., from the inverse self-diffusion constant) to a time scale associated with a non-diffusive process such as the single molecule rotational time, or the stress relaxation time, or the cage scale structural relaxation time deduced from the incoherent dynamic structure factor which is perhaps the most germane analogy in the context of our present discussion. Alternatively, “decoupling” has an analog for the problem of the failure of time-temperature superposition in cold polymer liquids where the left hand side refers to the ratio of a chain scale relaxation time to the local segmental relaxation time. Of course, in these bulk realizations, decoupling is a phenomenon fundamentally different than in thin films or near interfaces since it relates to the emergence of space-time dynamic heterogeneity in cold liquids via a mechanism that remain highly debated and not well understood \cite{10}. But, in essence, the common aspect of decoupling in bulk liquids and broken symmetry (interfaces) thin films is that the temperature dependence of different time scales become different in a manner that is mathematically captured by a formula such as eq 30. In the bulk, eq 30 applies to leading order with a material specific positive exponent less than unity \cite{10}, while near interfaces or in films the effective exponent obviously must depend on spatial location.

Finally, we emphasize that the double exponential form of the relaxation time gradient and the decoupling effect are not independent, and both are fundamentally tied to our prediction of barrier factorization with an exponential form of $f(z)$ per eq 23. Hence, one can write (to leading order) the following multiple inter-connections:
\begin{eqnarray}
\log\left(\frac{\tau_{\alpha,bulk}}{\tau_\alpha(z)}\right)&=&Ae^{-z/\xi}=\log\left(\frac{\tau_{\alpha,bulk}}{\tau_0^{*}}\right)\nonumber\\
&=&\varepsilon(z)\beta F_{total}^{bulk}(T)\nonumber\\
&=&\left(1-f(z)\right)\beta F_{total}^{bulk}(T)
\label{eq:31}
\end{eqnarray}
Thus, the amplitude and decay length in the double exponential can be expressed as:
\begin{eqnarray}
A(T)=\left(1-f(z=0)\right)\beta F_{total}^{bulk}(T)
\label{eq:32}
\end{eqnarray}
\begin{eqnarray}
e^{-z/\xi}=\frac{1-f(z)}{1-f(0)}=\frac{\varepsilon(z)}{\varepsilon(0)}
\label{eq:33}
\end{eqnarray}
The above relations clear show that \textit{both} the decoupling effect with an exponential in $z$ variation of the effective exponent and the double exponential form of the alpha time gradient follow from our core physical idea for the z-dependence of the caging part of the dynamic free energy which contains the idea of barrier factorization, as discussed in section III. 
\section{Summary and Discussion}
Based on our most recent work \cite{25} which extended the dynamic free energy concept of NLE theory to capture interface-nucleated mobility transfer effects in films, the present article has combined all of the bulk and thick film ECNLE-based theoretical ideas to establish how a sharp vapor interface induces spatial gradients of the collective elastic and total dynamic barriers and alpha relaxation time for the foundational hard sphere fluid and diverse polymer liquids. Our most fundamental prediction is twofold:  (i) the ratio of the total activation barrier at a distance $z$ from the vapor interface to its bulk analog is nearly independent of density or temperature (quasi-universal barrier factorization property), and (ii) the spatial variation of this ratio is nearly exponential with a penetration length that is largely temperature or density insensitive in the glassy dynamics regime. As a consequence, this provides a theoretical basis (the first to the best of our knowledge) for an alpha time gradient of a double exponential form to leading order, in qualitative accord with simulations. Hence, the double exponential behavior derives from our physical ideas concerning how kinetic constraints are encoded in the caging part of the dynamic free energy as a function of distance from the interface \cite{25}. By fitting the numerical theory results for the logarithm of the reduced alpha time to an exponential function, a decay length and amplitude (which quantifies the mobility acceleration at the interface) were extracted. While the range of the spatial decay is roughly constant ($\sim 3d$), the amplitude grows exponentially with volume fraction and is proportional to logarithm of the bulk alpha time to leading order. 

The temperature dependence of the alpha time strongly weakens upon approaching the vapor interface, and massively so for the top two layers which has strong implications for the formation mechanism of ultra-stable glasses \cite{57}. This behavior also results in large and relatively long range spatial gradients of the local $T_g$. Recovery of bulk behavior does not occur in a practical sense until $\sim 10-13$ particle diameters from the surface, which for typical organic molecules or polymers translates to $\sim 10-20$ nm. Our results for the alpha time and $T_g$ gradients suggest that in free standing films with two interfaces, strong interference or coupling effects likely emerges as films approach a thickness of order 20-40 nm. 

High resolution direct measurements of the form of the alpha relaxation time gradient near an interface is presently not possible. This partially motivated our analysis of three quantities that can (or potentially be) measured which are sensitive to the double exponential form of the alpha time gradient. First, we computed an average interfacial layer thickness of practical relevance to recent experiments. It grows roughly linearly with temperature as the film is cooled, and attains a rather large value at $T_g$ which is bigger for more fragile polymer liquids. Second, perhaps counter-intuitively, to leading order the ratio of the \textit{layer-averaged} interfacial relaxation time to its bulk is essentially universal-invariant to chemistry, volume fraction, temperature - and is of modest magnitude. Our analytic analysis reveals that this is a consequence of both how the dynamic gradient is averaged over in ensemble-averaged experiments \textit{and} (most importantly) the double exponential form of the relaxation time gradient with a nearly temperature invariant decay length. Third, we derived connections between the function $f(z)$ that quantifies the barrier factorization property (the exponential nature of which underlies the double exponential form of the alpha time gradient), and the functional form of the experimentally measureable glass transition temperature gradient. 

Power law dynamic decoupling of the $z$-dependent alpha time from its bulk value over a remarkably large range of time scales is predicted. Its existence is not fundamentally due to surface-nucleated dynamical constraint transfer since the more primitive ECNLE theory without the latter effect also makes this prediction \cite{44}. The reason is the generic nature of the barrier factorization property which follows from the ECNLE theory foundational concept of a spatially varying dynamic free energy near an interface. However, by including the new mobility transfer physics the degree of decoupling becomes much stronger, and with an effective exponent that now decays to zero exponentially with distance from the free surface, both in good agreement with simulation \cite{41}. This, in turn, reflects the most fundamental new prediction that the total activation barrier varies, to leading order, in an exponential manner as a function of distance from the vapor interface.
	
Finally, we note that the present work has only addressed the likely simplest case of a thick film with one flat vapor interface of negligible width. Whether or how our findings are relevant to other systems is unknown. But by building on the ideas reported here and in ref.\cite{25}, we are in the process of constructing new theories that include mobility transfer physics for thick films where the interface is solid (smooth or corrugated), and for finite thickness films where the presence of two identical or different interfaces must result in some level of non-additive interference effects when the films are thin enough. Our basic ideas are also not tied to planar films, and can be potentially extended to treat spherical or cylindrical geometries. The above extensions will be crucial for a host of puzzles (especially in experiment) that we have not addressed in the present article. An incomplete list includes: (i) the existence and/or nature of barrier factorization, double exponential alpha time gradients, and decoupling in these other more complex systems, (ii) non-monotonic dynamical effects as a function of film thickness \cite{33}, (iii) ultra-broad dynamical gradients and loss of any bulk region in thin enough films, (iv) possible two glass transitions in sufficiently thin free standing films of very high molecular weight polymers \cite{29} and asymmetric supported films \cite{30,31}, (v) the role of finite curvature of an interface or surface such as in droplets or polymer nanocomposites. 
\begin{acknowledgments}
This work was supported by DOE-BES under Grant No. DE-FG02-07ER46471 administered through the Materials Research Laboratory at UIUC. We thank Professor David Simmons for stimulating and informative discussions, and Dr. Yuxing Zhou for providing the simulation data from ref.\cite{37} and helpful discussions. 
\end{acknowledgments}
\section{Appendix A: Comparison of Theory and Simulation}
Here we comment on details of the simulations we have compared to in the main text. In the work of Hsu et al (H) \cite{51} and Zhou and Milner (ZM) \cite{37}, the local segmental dynamics was analyzed based on a layer-like model. The bulk 171-ns and 1-ns alpha relaxation times for PS in the latter simulations are roughly equivalent to a mapped hard sphere system in ECNLE theory at $\Phi=0.55$ and $\Phi=0.57$, respectively. In general, the various simulation results in the double logarithmic representation of $\tau_{\alpha,bulk}/\tau_\alpha(z)$ shown in Figure 4b exhibit features that are not quantitatively identical to the theory. This is not unexpected since the models simulated are not identical to hard spheres and both our coarse grained Kuhn sphere model and statistical mechanical theory are approximate. The mobility gradients found in the ZM \cite{37} and H \cite{51} simulations do agree well with our results near the interface where the cage-scale dynamics is significantly altered by the vapor surface. Farther from the interface (e.g., $z\geq 8d$) where the deviation of the alpha time in the film from its bulk analog is rather small, there are differences between theory and simulation, and also between the different simulations relative to each other. We suspect the main reasons for the former observation could be how one theoretically treats the collective elastic physics in films, and the varying importance of this aspect in different computational studies using different models at different thermodynamic states. 

A second order issue is the elementary length scale, $d$, that non-dimensionalizes distance from the vapor surface. For the theory this is the effective hard-sphere (Kuhn segment) diameter \cite{23}. To estimate its analog in simulations we used the experimental characteristic ratio for polystyrene (PS). The thermal mapping in ECNLE theory \cite{23,24} is based on the long chain limit where the characteristic ratio $C_{N=\infty}=9.5$. ZM \cite{37} used an atomistic model of PS with a degree of polymerization of $N = 10$.  Riggleman et al (RR) \cite{52} and Simmons et al (S) \cite{41} used a bead-spring model for glass-forming thin films and thick films, respectively. In the RR and S simulations, each chain consists of 20 beads that interact via a Lenard-Jones (LJ) potential. Using the known variation of the characteristic ratio with N for PS, we estimate $d_{ZM}=\left(3/9.5\right)^{1/3}d\approx 0.68d$ and $d_{RR}=\left(4.8/9.5\right)^{1/3}d\approx 0.8d$. Alternatively, for the RR and S simulations one could use $2.6\pi d_{RR}^3/6=\pi d^3/6$ or $d_{RR}=d/2.6^{1/3}\approx 0.73d$, which yields results very close to those stated above. For consistency, we employed $d_{RR}\approx d_{DS}\approx 0.8d$. For the H simulations \cite{51}, since their layer thickness is identical to ZM \cite{37}, we approximate $d_H=d_{ZM}$. This procedure is the basis of plotting the simulation data in terms of $z/d$ in Figure 4b. Finally, to compare quantitatively with experimental systems that employ real time, simulators often adopt the rough conversion that 1 LJ time unit approximately equals 1 ps \cite{60,61}, and we adopt this mapping in our comparisons.

\end{document}